\documentstyle[aps,floats,epsf]{revtex}

\begin{document}
\title{Lagrangian Tetrad Dynamics and the Phenomenology of Turbulence.}
\author{Michael Chertkov$^{1,3}$, Alain Pumir$^2$ and Boris I. Shraiman$^3$}
\address{$^1$ Department of Physics, Princeton University,Princeton, NJ 08544\\
$^2$ I. N. L. N., 1361 route des Lucioles, F-06560, Valbonne, France \\
$^3$ Bell Labs, Lucent Technologies,700 Mountain Ave, Murray Hill, NJ 07974}
\maketitle

\vspace{3cm}

\begin{abstract}
A new phenomenological model of turbulent fluctuations is constructed by
considering the Lagrangian dynamics of 4 points (the {\it tetrad}). The
closure of the equations of motion is achieved by postulating an {\it 
anisotropic}, i.e. tetrad shape dependent, relation of the local pressure
and the velocity gradient defined on the tetrad. The non-local contribution
to the pressure and the incoherent small scale fluctuations are modeled as
Gaussian white "noise". The resulting stochastic model for the
coarse-grained velocity gradient is analyzed approximately, yielding
predictions for the probability distribution functions of different 2nd and
3d order invariants. The results are compared with the direct numerical
simulation of the Navier-Stokes. 
The model provides a reasonable representation of the non-linear dynamics
involved in energy transfer and vortex stretching and allows to study
interesting aspects of the statistical geometry of turbulence, e.g.
vorticity/strain alignment. In a state with a constant energy flux (and K41
power spectrum) it exhibits the anomalous scaling of high moments associated
with formation of high gradient sheets - events associated with large energy
transfer. An approach to the more complete analysis of the stochastic model,
properly including the effect of fluctuations, is outlined and will enable
further quantitative juxtaposition of the model with the results of the DNS.
\end{abstract}

\draft



\section{Introduction}

The old problem of hydrodynamic turbulence has in recent years attracted
resurgent interest stimulated by the new generation of laboratory
experiments and the newly acquired ability of the direct numerical
simulations to probe interesting aspects of turbulence. In the light of the
new ideas and developments there has also been new appreciation of the
seminal contributions of Kolmogorov, reviewed in a recent book by Frisch 
\cite{UF}, and of Kraichnan to whom the present volume and this article are
dedicated. The key issues and the progress of the last years have been well
reviewed\cite{UF,SA} and are well represented in the present Festschrift.
Much effort has been dedicated to 1) documenting and understanding the
anomalous (i.e. non-Kolmogorov 41) scaling of high moments\cite
{UF,SA,AvA84,Tbl,MN,RHK90} associated with intermittency, 2) understanding the
structure and the local geometry of the intermittent regions of the flow\cite
{EDS81,AKKG,DCB,SJO91,VM,JIM93,JIM92,Tsi92,Cons}. On the theory side new ideas
derived from the new understanding of anomalous scaling of the Passive
Scalar problem \cite{RHK94,KYC95,GK,WG4,SS95} and of the Burger's turbulence\cite
{RHK90,Pol,Sin,Mez} both pioneered by Kraichnan\cite{RHK68}. Yet, the
theoretical description of turbulence based on first principles, i.e. on a
controlled approximation to the Navier-Stokes equations, is still over the
horizon and to proceed in the right direction one must rely on phenomenology.
One reason for pursuing
the modeling approach is the need to bridge the existing gulf between our
understanding of the scaling\cite{UF,SA} of turbulent fluctuations and their
structure or ''statistical geometry''\cite{Cons,Tsi97}. A step in this
direction will be the subject of the present paper.


Our goal here is to advance a phenomenological model for the probability
distribution function (PDF) of turbulent velocity fluctuations. We shall
start by noting that the longitudinal velocity difference between two
observation points\cite{MY}, while being most readily observable, seems a
poor candidate for a fundamental dynamical field in terms of which to
attempt a closed statistical description. The intuitive reason is that the
longitudinal velocity difference senses only one of the 8 locally
independent components of the velocity gradient tensor which govern the
dynamics of the velocity field. Instead we shall choose the fundamental
field to be the coarse grained velocity gradient tensor $M_{ab}\equiv \int_{
{\bf \Gamma }}d\vec{r}\partial _{a}v_{b}(\vec{r})$ defined over a region $
{\bf \Gamma }$ with characteristic scale $R$ lying in the inertial range. This
region may be best thought of as a local correlation volume of the velocity gradient
coarse-grained on scale $R$ - an "eddy" of sort.
The phenomenological model then will be based on the Lagrangian dynamics of
the ${\bf \Gamma }$-volume, parametrized by four points - the "tetrad" -
and its strain and vorticity fields as described
by $M_{ab}$. Effort will be made to preserve the essential non-linear
dynamics governing evolution of coarse grained strain and vorticity and the
concomitant distortion of the Lagrangian volume. This dynamics expresses the
fundamental constraints due to the conservation of energy and circulation.
In contrast, the dynamics of velocity fluctuations arising from the
scales smaller than that of the tetrad (and generating incoherent motion of the
points) will be
modeled as a Gaussian white process obeying K41 scaling. Essential element
of the theory advanced below will be the decomposition of the pressure into
the local part determined by the M-field via incompressibility and the
non-local part due to the contribution of distant regions, which will again
be modeled as a Gaussian random force. Such an approximation for the
pressure may perhaps be justifiable in large spatial dimension \cite{ft1,FFR,Yakh,WG4}, 
but we shall be content with exploring its consequences and comparing the results
with the direct numerical simulations.

We cannot hope to review here the evolution of the phenomenological modeling ideas,
yet we shall put the present work into the context of two recent efforts. 
The "PDF models" of Pope and coworkers\cite{Pope94} attempt to close the equation 
for the velocity probability distribution function (PDF) on the level 
of $one$ point: in contrast,
our model deals with relative velocity differences on 4 points which naturally brings in
Kolmogorov's ideas and allows to address the intermittency phenomenon. 
The two approaches however share the need to model the pressure Hessian/strain-
rate correlations (in our case on the coarse-grained level) and share the
realization that this model is improved by incorporating dynamical 
information about local
anisotropy\cite{Reynolds95,vSlP97}(in our case furnished by the moment 
of inertia tensor of the evolving Lagrangian volume). 
Another point of reference is the $two-point$ PDF closure advanced
by Yakhot\cite{Yakh} on the basis of the work of Kraichnan\cite{RHK94} 
and Polyakov\cite{Pol}. 
There too one arrived at a Fokker-Planck type equation for the PDF 
of velocity differences at given
point separation, yet the approach differs from the present one in the treatment 
of the correlations of large and small scale
fluctuations and our approach, by virtue of tracking a tetrad rather than a pair 
will retain more of the geometry of the flow.

The model will be presented in Section 2 in the form of the stochastic
equations of motion for two tensors specifying the coarse-grained velocity
gradient and the shape (i.e. moment of inertia) of the evolving Lagrangian
volume. We shall write down the corresponding Fokker-Planck equation for the
Probability Distribution Function and
discuss the energy transfer considerations which played the key role in the 
formulation of the model. Section 3 relates the deterministic aspects of
the model to the so called Restricted Euler (RE) dynamics that has been
investigated by Vieillefosse\cite{VFS} originally in the context of the
finite time singularities (see also L\'eorat\cite{Lrt}) and Cantwell and
coworkers \cite{C1,C2,C3} with the emphasis on the local topology of the
flow. RE describes the evolution of the velocity gradient at a point within
an isotropic approximation for the pressure which allows to close the Euler
equation locally. We shall see that elevation of the dynamics to the
coarse-grained level and the introduction of the second dynamical field to
keep track of the shape of the Lagrangian correlation volume (which depends on the $
history$ of the strain) allows to go beyond the isotropic pressure
approximation: the unphysical finite time singularity of RE is removed,
while the sensible short time dynamical properties (related for example to
the vorticity strain alignment \cite{AKKG}) are retained. Finally in Section
3 the deterministic dynamics will be compared with the empirical ''mean
field'' equation of motion for the coarse grained velocity tensor ${\bf M}$,
constructed from the conditional average 
$\left\langle\dot{{\bf M}}|{\bf M}\right\rangle$ measured
in the DNS of the Navier-Stokes at $R_{\lambda }=85$. In Section 4
we return to the stochastic model and  write down the formal solution of
the Fokker-Planck equation in terms of the path integral relating the probability
of a given coarse grained velocity gradient on a given inertial
range tetrad to the velocity PDF on the integral scale. This path integral
representation serves as a point of departure for the semiclassical approximation.
It also has a well defined
deterministic limit where the effect of the stochastic terms in the
tetrad dynamics can be neglected. In the latter limit the
probability of "observing" any given coarse-grained velocity gradient 
on an inertial range 
tetrad is determined by the probability of its integral scale preimage and
can be calculated by integrating the equations of motion backward in time.
This crude but simple approximation is employed in Section 5 in order to
gain insight into the behavior of the model and to
elaborate its statistical predictions emphasizing energy
transfer, enstrophy and its production and the alignment of vorticity and
strain. The comparison of the results with the direct numerical simulation
of the Navier-Stokes equations is quite encouraging. The calculated
probability distribution function also exhibits anomalous scaling of high
moments. In conclusion, Section 6 is a summary and the outline of further
inquiry.

\section{The Minimal Model}

The minimal parametrization of the ${\bf \Gamma }$ volume is a tetrahedron (more
generally a $d$-dimensional simplex) defined by four, hence {\it tetrad},
(or $d+1$) Lagrangian points, $\vec{r}_{\alpha }(t)$, which upon elimination
of the center of mass define a triad of vectors $\vec{\rho}_{i}$: 
$\vec{\rho}_{1}=(\vec{r}_{1}-\vec{r}_{2})/\sqrt{{2}}$, 
$\vec{\rho}_{2}=(\vec{r}_{1}+\vec{r}_{2}-2\vec{r}_{3})/\sqrt{{6}}$, 
$\vec{\rho}_{3}=(\vec{r}_{1}+\vec{r}_{2}+\vec{r}_{3}-3\vec{r}_{4})/\sqrt{{12}}$. 
It will be useful to treat this
triad of vectors as $3\times 3$ matrix, $\rho _{i}^{a}$, where $a$ is the
spatial index. Analogously, by eliminating the center of mass velocity from
the instantaneous velocity of the vertices, $\dot{r}_{\alpha }$, one can
define a triad of relative velocities, $\vec{v}_{i}$. The coarse grained
gradient field can now be defined simply by interpolation: 
\begin{equation}
M_{ab}=(\rho ^{-1})_{i}^{a}v_{i}^{b}-{\frac{\delta _{ab}}{3}}tr\,(\rho
^{-1}v).  \label{1}
\end{equation}

Alternatively and more generally one may decompose the ''observed'' velocity
differences, $v_{i}^{a}$ into a slow component arising from the scales
greater than the radius of gyration, ${\bf \geq }R$, represented by the
coarse grained velocity gradient matrix $M_{ab}$ and the rapidly fluctuating
incoherent component, $u_{j}^{a}$ arising from scales $\ll R$.
\begin{equation}
v_{i}^{a}\,=\,\rho _{i}^{b}M_{ab}+u_{i}^{a}.  \label{2}
\end{equation}
The strategy will be to derive the dynamics of $M_{ab}$ and $\rho _i^{a}$
while treating ${\bf u}$ as a Gaussian white noise with the statistics
depending on the Kolmogorov's energy dissipation rate $\epsilon $ as well as
the instantaneous ${\bf \rho }$ and ${\bf M}$. Lagrangian dynamics is
governed by ${\frac{D}{Dt}}{\bf v}_{i}=-{\bf \nabla }p_{i}\,+\,{\bf f}_{i}$
and ${\frac{D}{Dt}}{\bf \rho }_{i}={\bf v}_{i}$, where ${\bf f}_{i}$ is the
random external force and $\nabla p_{i}$ is the pressure gradient, both
properly differenced between the observation points. The dynamics of the
coarse grained velocity gradient tensor, ${\bf M}$, and the tetrad tensor $%
\rho $ has the form: 
\begin{mathletters}
\begin{eqnarray}
{\frac{d}{{dt}}}M_{ab}\,+\,M_{ab}^{2}-\Pi _{ab}trM^{2}\, &=&\,\xi _{ab},
\label{3a} \\
{\frac{d}{{dt}}}\rho _{i}^{a}\,-\,\rho _{i}^{b}M_{ba}\, &=&\,u_{i}^{a},
\label{3b} \\
\Pi _{ab} &\equiv &k_{i}^{a}k_{i}^{b}/tr{\bf kk}^{\dagger },  \label{3c}
\end{eqnarray}
with matrix ${\bf k}\equiv {\bf \rho }^{-1}$. The left hand sides of (\ref
{3a},\ref{3b}) describe the self-advection and stretching of the tetrad by
the coherent (on the scale of $\rho $ ) component of the velocity field. The
right hand side of (\ref{3a}) derives from the pressure gradient and the
random force differences as well as from the coupling to incoherent small
scale fluctuations. The $\Pi _{ab}\,trM^{2}$ term, with $tr{\bf \Pi }=1$ on
the left hand side represents the ''local'' component of the pressure needed
to insure conservation of $tr{\bf M}=0$ as required by incompressibility.
Tensor $\Pi _{ab}$ is a measure of tetrad anisotropy representing the anisotropy
of the Lagrangian correlation volume built up by prior evolution \cite{ft2}. 
This choice of the
local term, in contrast with the simpler, isotropic form \cite{VFS} is
dictated by the requirement that the pressure forces should do no work and
drop out of the energy balance (see below). In addition it reintroduces
proper reduction of the deterministic dynamics (left hand side of (\ref{3a}
)) to 2D: i.e. a 2D velocity gradient configuration remains 2D provided that
the tetrad has the shape of a filament, i.e. ${\bf \rho }^{\dagger }{\bf %
\rho }$ is a tensor of rank one. The remaining non-local part of the
pressure is subsumed in $\xi _{ab}$. 

We now define the stochastic components of the model
appearing on the right hand sides of (3a-b).  It is appealing to model the non-local
part of the pressure retained in $\xi $, along with the contribution of
small scales, as $\delta $-correlated Gaussian random noise with the
variance depending on the energy flux $\epsilon $ as well as the
instantaneous ${\bf \rho }$ and ${\bf M}$. Let us consider a polynomial :
\end{mathletters}
\begin{equation}
{\bf \xi }={\bf \eta }+{\bf \zeta M}+\alpha ({\bf M}^{2}-\Pi \,tr{\bf M}%
^{2}).  \label{4}
\end{equation}
where $\eta $ is a random matrix, and $\zeta $ a random function. The
dimension of both $\xi $ and ${\bf M}^{2}$ is time$^{-2}$ so that $\alpha $
is a constant. The last term is clearly not the most general one can write.
It is however the one suggested by the numerical study of the statistics of
the right hand side of (\ref{3a}), originally by Borue and Orszag \cite{BO98}. 
According to the DNS the average $\xi _{ab}$ conditioned on ${\bf M}$ is
not zero, but is reasonably well approximated by $\alpha ({\bf M}^{2}-{\bf
\Pi }tr{\bf M}^{2})$ with $0<\alpha <.8$, depending on the scale, at least
for the isotropic tetrad (i.e. ${\bf \Pi =1}/3)$. We shall assume that in
the inertial range $\alpha $ is constant (which corresponds to keeping only the 
deterministic component of the 3d term in (4)) and take it to be a model
parameter. This $\alpha $ term ''renormalizes'' the time scale of the
deterministic dynamics described by the left hand side of the equation (\ref
{3a}) and will have an important consequence for the energy transfer in the
model, as we shall see shortly.

Let 
\begin{mathletters}
\begin{equation}
\left\langle \eta _{ab}(t)\eta _{cd}(0)\right\rangle ={\frac{2C_{\eta
}\epsilon }{tr\rho \rho ^{\dagger }}}\,[\delta _{ac}\delta _{bd}-{\frac{1}{3}%
}\delta _{ab}\delta _{cd}]\,\delta (t)  \label{5a}
\end{equation}
which is the simplest form respecting incompressibility, with Kolmogorov's
energy flux $\epsilon $ and $C_{\eta }$ - a dimensionless parameter. Random $%
\eta $ causes diffusion in velocity space; note that $\epsilon $ has the
dimension of corresponding diffusivity. 
The appearence of $\epsilon$ in (5a) is further supported by the fact that $\eta$ 
fluctuations contribute to the energy transfer as we shall see below.
In the ''minimal'' model, which we
are now constructing we shall drop the possible multiplicative random field $%
\zeta $ (see Eq.(\ref{4})).

The fluctuations of the small scale\cite{ft3} ${\bf u}$ can be resolved into parts
which are longitudinal and transverse to $\rho _{i}^{a}$: 
\begin{equation}
\left\langle u_{i}^{a}(t)u_{j}^{b}(0)\right\rangle \,=\,2C_{||}\sqrt{tr{{\bf M}%
{\bf M}^{\dagger }}}\,\rho _{i}^{a}\rho _{j}^{b}\,\delta (t)+2C_{\perp }%
\sqrt{tr{{\bf M}{\bf M}^{\dagger }}}\,(\rho ^{2}\delta ^{ab}\delta
_{ij}\,-\,\rho _{i}^{a}\rho _{j}^{b})\,\delta (t)  \label{5b}
\end{equation}
where with the K41 scaling in mind we take the characteristic time to be the
''eddy turnover'' time, $1/\sqrt{tr{{\bf M}{\bf M}^{\dagger }}}$. The
longitudinal part of ${\bf u}$ in the $\rho $-equation (\ref{3b}) would by
itself produce Richardson diffusion behavior, $\left\langle \rho
^{2}(t)\right\rangle \sim \epsilon t^{3}$ provided that the Kolmogorov
scaling $\sqrt{tr{{\bf M}{\bf M}^{\dagger }}\,}{\bf \sim }\epsilon ^{1/3}\rho
^{-2/3}$ holds. However, Richardson diffusion would also arise from the
non-Gaussian coherent stretching term $\rho {\bf M}$, and the Gaussian
longitudinal fluctuations, $C_{||}$, do not appear to be essential. We shall
set $C_{||}=0$. The transverse fluctuations $C_{\perp }$ however are very
important, because in their absence the effect of volume preserving coherent
stretching would lead to the rapid growth of anisotropy of the tetrad. The
incoherent transverse velocity fluctuations act to redistribute the vertices
of the tetrad uniformly on the surface of the $\rho ^{2}\equiv tr{\bf \rho }%
^{\dagger }{\bf \rho }$ hyper-sphere in $d^{2}=9$ dimensions, thus
introducing the isotropisation mechanism. The competition of the coherent
stretching which leads to the growth of the radius of gyration $\rho ^{2}$
(both forward {\it and backward} in time) and the isotropisation over the $%
\rho ^{2}$ shell will play the key role in setting up the energy flux.

The stochastic tetrad dynamics defined by Eq. (3-5) 
determines the Lagrangian transition probability
from tetrad $({\bf M}^{\prime },{\bf \rho }^{\prime })$ to $({\bf M},{\bf %
\rho })$ at a time $t$ later, $G_{t}({\bf M},{\bf \rho }|{\bf M}^{\prime },%
{\bf \rho }^{\prime })$, which satisfies a Fokker-Planck equation
\end{mathletters}
\begin{equation}
\left(
{\frac{\partial }{\partial t}}-{\bf L}\right)
\,G_{t}({\bf M},{\bf \rho }|{\bf M}%
^{\prime },{\bf \rho }^{\prime })\,=\,\delta ({\bf M}-{\bf M}^{\prime
})\,\delta ({\bf \rho }-{\bf \rho }^{\prime })  \label{6}
\end{equation}
with the evolution operator
\begin{eqnarray}
{\bf L} &=&(1-\alpha ){\frac{\partial }{\partial M_{ab}}}
\left(M_{ab}^{2}\,-\,
\Pi _{ab}\,tr{\bf M}^{2}\right)
-{\frac{\partial }{\partial \rho _{b}^{i}}}\,\rho
_{a}^{i}M_{ab}  \label{7} \\
&+&C_{\eta }{\frac{\epsilon }{{\rho ^{2}}}}
\left({\frac{\partial ^{2}}{{\partial
M_{ab}\partial M_{ab}}}}-{\frac{1}{3}}{\frac{\partial ^{2}}{{\partial
M_{aa}\partial M_{bb}}}}\right)+
C_{\perp }\sqrt{tr{{\bf M}{\bf M}^{\dagger }}}{\frac{%
\partial }{\partial \rho _{i}^{a}}}\,(\rho ^{2}\delta ^{ab}\delta
_{ij}\,-\,\rho _{i}^{a}\rho _{j}^{b})\,{\frac{\partial }{\partial \rho
_{j}^{b}}}.  \nonumber
\end{eqnarray}

The invariant joint distribution, $P({\bf M},{\bf \rho })$, satisfying 
\begin{equation}
\partial _{t}P({\bf M},{\bf \rho })\,=\,{\bf L}P({\bf M},{\bf \rho })\,=\,0
\label{8}
\end{equation}
can be interpreted as the Eulerian PDF $P({\bf M},{\bf \rho })$ provided
that the normalization $\int d{\bf M}\,P({\bf M},{\bf \rho })=1$ is imposed
\cite{ft4}.
Equations (7-8) once supplemented
with the boundary condition specifying the Eulerian PDF on the integral scale,
$\rho^2 = L^2$, completely define our model.

Before proceeding with the analysis of the Fokker-Planck equation (8),
let us examine the energy balance, which was one of the key consideration in the
formulation of the model:
\begin{eqnarray}
{\frac{1}{2}}\partial _{t}<tr{\bf VV}^{\dagger }>_{\rho } &\equiv &\int d
{\bf M}\,\,tr({\bf \rho M}{\bf M}^{\dagger }\rho ^{\dagger })\,\,\partial
_{t}P({\bf M},{\bf \rho })  \nonumber \\
&=&-{\frac{\partial }{{\partial \rho _{a}^{i}}}}<V_{a}^{i}\;tr({\bf VV}%
^{\dagger })>_{\rho }+\,\alpha <tr({\bf VV}^{\dagger }{\bf M})>_{\rho } 
\nonumber \\
&+&{\frac{16}{3}}\,C_{\eta }\epsilon -C_{\perp }D_{d}\,+\,C_{\perp }f,
\label{9}
\end{eqnarray}
which is obtained by multiplying eq (\ref{8}) by $tr{\bf VV}^{\dagger }$
(where $V_{a}^{i}\equiv \rho _{b}^{i}M_{ba}$) and averaging with respect to $%
{\bf M}$. Note that the average $<...>_{\rho }$ is taken at fixed $\rho $
and remains a function of it. The first two terms on the right hand side may
be identified as the divergence of the large scale energy flux and the eddy
damping respectively.

Note that the term originating from the deterministic component of the
pressure $\Pi _{ab}\,tr{\bf M}^{2}$ drops out: the particular form of $\Pi
_{ab}$ was chosen for that purpose on the grounds that the pressure
gradients should not contribute to the energy transfer as seen in the von
K\'{a}rm\'{a}n-Howarth derivation \cite{LL}. However, since $V_{a}^{i}$ is only the
coarse grained and not the full local velocity, in contrast with the von
K\'{a}rm\'{a}n - Howarth analysis the divergence of the energy flux is balanced not
directly by the viscous dissipation term, but by the eddy damping. There are
also additional contributions due to the coupling with small scale
fluctuations represented by the last three terms in (\ref{9}). The $C_{\eta
} $ term represents the {\it diffusive} component of the energy flux arising
from the small scale fluctuations and the coupling of the tetrad to the
neighboring regions (entering via Gaussian $\eta $). The $C_{\perp }f$ term
represents the transverse energy flux with:
\begin{equation}
f\,=\,{\frac{\partial }{\partial \rho _{i}^{a}}}(\rho ^{2}\delta ^{ab}\delta
_{ij}\,-\,\rho _{i}^{a}\rho _{j}^{b})\,\left[ \,{\frac{\partial }{\partial
\rho _{j}^{b}}}\,\left\langle \sqrt{tr{{\bf M}{\bf M}^{\dagger }}}\,tr{\bf VV}%
^{\dagger }\right\rangle _{\rho }-2\rho _{j}^{c}\left\langle \sqrt{tr{{\bf M}%
{\bf M}^{\dagger }}\,}({\bf M}{\bf M}^{\dagger })_{bc}\right\rangle _{\rho
}\,\right]  \label{10}
\end{equation}
which redistributes the energy within the $\rho ^{2}=const$ shell, while the 
$C_{\perp }D_{d}$ is the diffusive contribution to the eddy damping: 
\begin{mathletters}
\begin{equation}
D_{d}\,=\,18\,\,\left\langle \sqrt{tr{{\bf M}{\bf M}^{\dagger }}}\,tr\,[(\,%
{\bf \rho }^{\dagger }{\bf \rho }-{\frac{{\bf 1}}{3}}\rho ^{2})\,{\bf M}{\bf %
M}^{\dagger }]\right\rangle \,_{\rho }.  \label{11a}
\end{equation}
This $D_{d}$ is reminiscent of the Smagorinsky\cite{RE,BO98} form of eddy
damping (popular in sub-grid simulations \cite{RE,Men}) but with significant
difference that in the latter the $d$-wave projector $(\,{\bf \rho }%
^{\dagger }{\bf \rho }-{\frac{{\bf 1}}{3}}\rho ^{2})\,$ appearing in (\ref
{11a}) is replaced by the simple scale factor $\rho ^{2}$. Hence, in
contrast with the Smagorinsky model, our diffusive damping term is only
active to the extent that the $\left\langle \sqrt{tr{{\bf M}{\bf M}^{\dagger }}%
}\,({\bf M}{\bf M}^{\dagger })_{ab}\right\rangle \,_{\rho }$ tensor is
correlated with $({\bf \rho }^{\dagger }{\bf \rho })_{ab}$ tensor within the 
$\rho ^{2}$ shell. Strictly speaking $D_{d}$ is not positive definite and
its interpretation as the {\it damping} term is contingent on the
expectation that the tetrad dynamics builds up the alignment of the
principal axis of ${\bf \rho }^{\dagger }{\bf \rho }$ and ${\bf M}{\bf M}%
^{\dagger }$.

Notably, the deterministic eddy damping term which has appeared in (\ref{9}) 
\begin{equation}
D_{nl}\,=\,-\,\alpha <tr{\bf V}^{\dagger }{\bf VM}>_{\rho }  \label{11b}
\end{equation}
is a direct generalization of the so-called {\it non-linear} eddy damping $%
\rho ^{2}\,<tr\,{\bf M}^{2}{\bf M}^{\dagger }>$ advanced by Bardina et al.%
\cite{Bard} and reduces to it for isotropic tetrads ${\bf \rho \rho }%
^{\dagger }={\bf 1}\rho ^{2}$. In this limit $D_{nl}\rightarrow -\,tr({\bf s}%
^{3}\,-\,{\bf \Omega \cdot s\cdot \Omega }\,)$, where ${\bf s}$ and ${\bf %
\Omega }$ are respectively the symmetric and antisymmetric parts of ${\bf M}$%
. Thus the energy transfer down scale is due to negative strain skewness or
positive enstrophy production \cite{MY,UF} (i.e. vortex stretching). We can
define the energy flux by averaging over the fixed $\rho ^{2}$ shells. Let $%
R\equiv \sqrt{{\rho ^{2}}}$ and $V_{R}\equiv \rho _{i}^{a}V_{i}^{a}/R$
denote the longitudinal velocity, then 
\end{mathletters}
\begin{equation}
\epsilon \,=\,-\partial _{R}\left\langle {\bf V}_{R}\,tr{\bf VV}^{\dagger
}\right\rangle _{R}+{\frac{16}{3}}\,C_{\eta }\epsilon  \label{12}
\end{equation}
is balanced by eddy damping $\epsilon \,=\,D_{nl}+C_{\perp }D_{d}$. Here $%
\left\langle ...\right\rangle _{R}$ denotes an additional average over $\rho
^{2}=R^{2}$ shell.

Below we will often think of the diffusive contributions as being small
compared to the non-linear interactions on current scale: that is, we shall
assume $C_{\eta },\,C_{\perp }\ll 1$ and treat them as (a singular)
perturbation of the deterministic dynamics. Another tractable and perhaps
physically more plausible limit is $C_{\perp }\gg 1$.

\section{ Deterministic Dynamics and the Restricted Euler Model.}
Note that the equation of the form (\ref{3a}) also governs the Lagrangian
evolution of the actual $local$ velocity gradient matrix $m_{ab}=\partial
_{a}v_{b}$ (we use lower case ${\bf m}$ to avoid confusion with the coarse
grained object) 
\begin{mathletters}
\begin{equation}
{\frac{d}{{dt}}}m_{ab}\,+\,m_{ab}^{2}\,=\,-\partial _{a}\partial _{b}p
\label{13a}
\end{equation}
as derives from the Euler equation. L\'eorat \cite{Lrt} and Vieillefosse \cite
{VFS} have considered (\ref{13a}) retaining only the {\it local} and {\it %
isotropic} contribution to the pressure 
\begin{equation}
\partial _{a}\partial _{b}\,p\,=\,-\,{\frac{\delta _{ab}}{3}}\,tr\,{\bf m}%
^{2}  \label{13b}
\end{equation}
as a model of vorticity dynamics and observed that (\ref{13a},\ref{13b})
leads to a finite time singularity with $||{\bf m|}|{\bf \sim }%
(t_{*}-t)^{-1} $. The dynamics governed by (\ref{13a},\ref{13b}) - the
''Restricted Euler dynamics'', to use Cantwell's terminology \cite{C1} 
lies entirely in the two-dimensional phase space defined by the two
invariants \cite{VFS,C1} $tr\,{\bf m}^{2}$ and $tr\,{\bf m}^{3}$. This
reduction stems from the $SL(3)$ invariance, ${\bf m\rightarrow g\,m\,g}%
^{-1} $ with ${\bf g}$ being an arbitrary $3\times 3$ matrix, which allows
one to bring ${\bf m}(t)$ to diagonal form ${\bf \Lambda }(t)$ by a {\it %
time independent} similarity transformation ${\bf m}(t)={\bf U\Lambda }(t)%
{\bf U}^{-1}$. There is yet one more independent constant of motion found by
Viellefosse\cite{VFS}: the ''discriminant'' $D\equiv 3(tr\,({\bf m}
^{3}))^{2}-{\frac{1}{2}}(tr\,({\bf m}^{2}))^{3}\,=\,%
-(\lambda _{1}-\lambda _{2})^{2}(\lambda _{2}-\lambda _{3})^{2}(\lambda
_{3}-\lambda _{1})^{2}$, where $\lambda _{i}(t)$ are the (in general
complex) eigenvalues of ${\bf m}(t)$. The RE dynamics thus reduces to 1D
flow, i.e. 
it is integrable! Figure 2 shows the flow in the 2D phase plane of the
Cantwell's invariants \cite{C1} $Q\equiv -{\frac{1}{2}}\,tr\,\,{\bf m}^{2}$, $%
R\equiv -{\frac{1}{3}}\,tr\,{\bf m}^{3}$ and the finite time singularity
corresponds to the $R\rightarrow \infty $, $Q\rightarrow -\infty $
asymptotically approaching the $D=0$ separatrix.

Along the $D=0$ separatrix the flow is particularly simple: 
\end{mathletters}
\begin{equation}
{\bf m}(t)=\left( 
\begin{array}{lll}
\lambda (t) & 0 & 0 \\ 
0 & \lambda (t) & 0 \\ 
0 & 0 & -2\lambda (t)
\end{array}
\right)  \label{14}
\end{equation}
with $\lambda (t)=\lambda (0)/(1-t\lambda (0))$ making the finite time
singularity at $t_{*}=1/\lambda (0)$ explicit.

The region above the separatrix, $D>0$, is elliptic: the eigenvalues of the
velocity gradient eigenvalues become complex and the Lagrangian trajectories
are rotating; the region below the separatrix, $D<0$, is hyperbolic: the
eigenvalues are real and the trajectories are strain dominated. These
topological aspects of RE dynamics were emphasized by Blackburn et al\cite
{C3}.

As a model of finite time singularity RE solutions were rejected\cite{ft5}
on the reasonable ground that if considered as global solutions of Euler equations these do not
satisfy sensible boundary conditions and have unbounded energy. 
Ashurst et
al\cite{AKKG} however noted that the statistics of the vorticity/strain
alignment observed in the DNS of Navier-Stokes may be qualitatively
understood in terms of RE. Subsequently Cantwell and coworkers \cite
{C1,C2,C3} proceeded to investigate the DNS generated statistics of $R,Q$
invariants and observed that the probability distribution function (PDF) of $%
R,Q$ exhibits a pronounced tail along the Viellefosse $D=0$ asymptote as can
be seen on Fig.3. These two observations suggest that despite the draconian
local and isotropic approximation to pressure and the unphysical finite time
singularity the RE dynamics does capture certain statistical features of the
physical flow.

The deterministic part of the Lagrangian tetrad dynamics defined in Section
2 generalizes RE by reinterpreting the velocity gradient tensor as a
coarse-grained field defined over the tetrad ${\bf \rho }$ and completing
the Lagrangian picture by adding the dynamical equation for ${\bf \rho }(t)$%
. The ${\bf \rho }$ field introduces the measure of current length scale and
the dependence on the history of the strain which controls the ''shape'' of
the tetrad. The ${\bf \rho }$-dynamics (\ref{3b}) is coupled to ${\bf M}$%
{\bf \ }via the anisotropy tensor ${\bf \Pi }$. For an isotropic tetrad
(i.e. regular tetrahedron) $\Pi _{ab}=\delta _{ab}/3$ and the $M$-dynamics
(the left hand side of (\ref{3a})) reduces instantaneously to RE equation (%
\ref{13a},\ref{13b}). In the next instant however the tetrad will become
distorted through the action of the volume preserving ${\bf M}$ following
the $\rho $-dynamics equation (left hand side of (\ref{3b})) and the
trajectory will come out of the RE plane. Its evolution will depart from RE
as the anisotropy increases and at some point the growth of $||{\bf M}||$
will be cut off. This is most easily seen for the $D=0$ Vieillefosse line.
The dynamics of $\lambda $ (see (\ref{14})) becomes ${\dot{\lambda}}%
\,=\,(6q^{-1}-1)\,\lambda ^{2}$, with ${\bf \Pi =}diag\{1,1,q-2\}/q$ where $%
q $ evolves according to ${\dot{q}}\,=6(q-2)\lambda $. The isotropic tetrad
corresponds to $q=3$. Starting from isotropy and $\lambda >0$ both $\lambda
(t)$ and $q(t)$ grow. The growth of $q$ corresponds to the contraction of
one of the principle axis of the ${\bf \rho }^{\dagger }{\bf \rho }$ tensor
as the tetrad is flattened in a pancake. However when $q>6$ the growth of $%
\lambda $ reverses. Thus, anisotropy caused by the stretching of the tetrad
cuts off the Viellefosse finite time singularity\cite{ft6}. The modified RE dynamics however
retains the initial growth of ${\bf M}$ with two expanding and one
contracting strain directions and the consequent deformation of the tetrad
into a pancake or ribbon.\cite{ft7}
This process is the fundamental step of energy transfer\cite{MY,Betch}. In
the next Section we will see that the Viellefosse tail (large $R>0,Q<0$
region) of the Cantwell PDF on Fig.3 which is generated through this
process, indeed corresponds to large negative strain skewness associated
with the energy transfer \cite{MY}. Another retained aspect of the RE
dynamics is the evolution of the vorticity/strain alignment from
configurations where vorticity is parallel to $\alpha $-strain (i.e. the
fast stretching direction) to configuration where vorticity is aligned with
the intermediate, $\beta $-strain as observed numerically\cite
{EDS81,Kerr,AKKG}. The new feature of the modified model is that while in
the isotropic RE all 2D configurations of ${\bf M}$ evolve into 3D, (e.g. $%
M_{ab}\,=\,\epsilon _{abz}\,\omega _{z}$ will in the next instant acquire,
due to low local pressure, a contracting component of strain acting along the $z$%
-direction which will act to destroy $\omega _{z}$), the new anisotropic
model allows the $2D$ configurations of ${\bf M}$ to persist provided that the $%
{\bf \Pi }$ tensor is rank two (${\bf \rho }^{\dagger }{\bf \rho }$ rank
one) which correspond to filament like tetrads. Note that both intense
vorticity and quasi-one dimensional tetrads will be produced by the action
of strain with one stretching and two contracting directions $tr[{\bf s}%
^{3}]>0$, thus there potentially is a chance of describing vortex ''worms''%
\cite{EDS81,DCB,VM,JIM93}. We shall return to the discussion of the
kinematics of energy transfer and vortex stretching in Section 5.

How can one compare the deterministic tetrad dynamics model with the real
Navier-Stokes dynamics ? The relevant empirical object is the average $d{\bf %
M}/dt$ and $d{\bf \rho }/dt$ conditioned on ${\bf M}$, ${\bf \rho }$ but to
simplify matters we will restrict to isotropic tetrads and examine the flow
in the $Q,R$ phase space generated by the conditional averages $\left\langle 
{\dot{R}}|R,Q\right\rangle $ and $\left\langle \dot{Q}|R,Q\right\rangle $.
The latter were obtained by a DNS of Navier-Stokes.

Briefly, the Navier-Stokes equations are integrated by a standard
pseudo-spectral algorithm. Our code is fully de-aliased. We used up to $%
(128)^{3}$ collocation grid points, and the effective resolution was
maintained to be higher than $k_{\max }\eta \geq 1.4$, where $k_{max}$ is
the highest wave-vector in the simulation, and $\eta _{K}$ the Kolmogorov
length scale ($\eta _{K}\equiv (\nu ^{3}/\epsilon )^{1/4}$). Statistics were
accumulated for at least $3$ eddy turnover times. In the following, we
present our results for a Taylor scale $Re_{\lambda }=85$. Our investigation
of the influence of the Reynolds number in the range $20\le Re_{\lambda }\le
85$ did not reveal any major qualitative change of the statistics presented
here.

Fig. 4a,b,c show the streamlines in the $(R,Q)$ plane, reconstructed from the
conditional averages of $\left\langle {\dot{R}}|R,Q\right\rangle $ and $%
\left\langle \dot{Q}|R,Q\right\rangle $ computed numerically for three
different $\rho ^{2}$. The latter were increasing from the dissipation range
to large scale. For isotropic ${\bf \rho }$ our ${\bf M}$-dynamics is
instantaneously tangent to the RE and therefore the empirical flows can be
compared with Fig. 2. Remarkably, while there are significant deviations in
the topology of the flow for $\rho $ in the dissipative range (Fig.4a), the
instantaneous flow for large scale $\rho $ is surprisingly close to RE. The
deviations at small scales are presumably due to the viscous effects. The
conditional flow for $|\rho |\geq 10\eta _{k}$ can be fitted by the modified
RE: 
\begin{equation}
\frac{d{\bf M}}{dt}=(\alpha -1)({\bf M}^{2}-{\bf \Pi }Tr{\bf M}^{2})
\label{15}
\end{equation}
with $\alpha $ decreasing with increasing $|\rho |/\eta _{K}$ from .8 to 0.
For the reasons related to the energy transfer, discussed in Section 2, we
believe that $\alpha $ should be constant in the inertial range. The
continuous scale dependence observed in the fit to DNS however is not
unexpected, because the inertial range at the accessible $Re$ is quite
limited. On the other hand the approximate validity of (\ref{15}) as a
description of the coarse-grained Lagrangian evolution is quite encouraging.
It would be important to extend the comparison of the deterministic dynamics
(\ref{15}) with the numerical simulation for anisotropic tetrads, however in
that case the SL(3) invariance of (\ref{15}) is lost and in addition to the $%
R,Q$ invariants the time derivative must be conditioned on the vorticity,
which makes the computation more demanding statistically. It would also be
important to investigate systematically the deviations of the conditional
flow from (\ref{15}): these are expected to arise from the possible
additional deterministic terms in (\ref{15}) (e.g. $\gamma {\bf M}$) as well
as the stochastic dynamics. Much further work is required in this direction.

\section{Lagrangian Path Integrals and the Semiclassical Approximation.}
Let us now explore the statistical properties of the coarse-grained $M$%
-field on the tetrad $\rho $. The probability distribution $P({\bf M},\rho )$
is governed by the the Fokker-Planck equation (\ref{7},\ref{8}) but requires
specification of an additional boundary condition. Since the PDF of velocity
is known to be Gaussian on the integral scale we shall fix \cite{ft8}.
\begin{equation}
P({\bf M,\rho })|_{\rho ^{2}=L^{2}}\,\sim \exp \left[ -{\frac{tr{\bf M}{\bf M%
}^{\dagger }}{(\epsilon L^{-2})^{2/3}}}\right] \,.  \label{16}
\end{equation}

To impose the integral scale boundary condition one may use a generalization
of Green's theorem
\begin{eqnarray}
P({\bf M}^{\prime },{\bf \rho }^{\prime }{\bf )} &=&\int d{\bf M}\int d{\bf %
\rho }\left[ P({\bf M},{\bf \rho }){\bf L}^{\dagger }{\bf g}^{\dagger }({\bf %
M},{\bf \rho }|{\bf M}^{\prime },{\bf \rho }^{\prime })\,-\,{\bf L}P({\bf M},%
{\bf \rho }){\bf g}^{\dagger }({\bf M},{\bf \rho }|{\bf M}^{\prime },{\bf \rho }%
^{\prime })\right]  \nonumber \\
&=&\int d{\bf M}\,\int_{\rho ^{2}=L^{2}}tr(\,d{\bf \rho } \,{\bf M}^{\dagger
}{\bf \rho }^{\dagger })\,\,P({\bf M},{\bf \rho })\,{\bf g}^{\dagger }({\bf M}%
,{\bf \rho }|{\bf M}^{\prime },{\bf \rho }^{\prime })  \label{17}
\end{eqnarray}
where ${\bf L}^{\dagger }$ denotes the adjoint operator which governs
evolution backward in time (obtained by ${\bf M}\rightarrow -{\bf M}$) and $%
{\bf g}^{\dagger }{\bf (M,\rho |M}^{\prime }{\bf ,\rho }^{\prime }{\bf )}%
\,=\,{\bf L}^{\dagger -1}$ its static Green function.


The static Green's function ${\bf g}^{\dagger }$ is computed via the
Lagrangian Green's function (\ref{6}):
\begin{equation}
{\bf g}^{\dagger }{\bf (M,\rho |M}^{\prime }{\bf \rho }^{\prime }{\bf )\,}%
=\,\int_{-\infty }^{0}dT\,G_{T}({\bf M,\rho |M}^{\prime }{\bf \rho }^{\prime
})  \label{18}
\end{equation}
which has an intuitively appealing path integral representation:
\begin{equation}
G_{-T}({\bf M,\rho |M}^{\prime }{\bf \rho }^{\prime })\,=\,\int_{{\bf M}(-T)=%
{\bf M}^{\prime }}^{M(0)={\bf M}}\,D{\bf M}\,\int_{\rho (-T)=\rho ^{\prime
}}^{\rho (0)=\rho }\,D\rho \,\ exp[-S(\{{\bf M},{\bf \rho }\})]  \label{19}
\end{equation}
summing over all possible paths connecting initial ${\bf M}^{\prime },\,{\bf %
\rho }^{\prime }${\bf \ }at time $-T$ with the final ${\bf M,\,\rho }$ at
time 0 weighted with the action 
\begin{eqnarray}
S &=&\frac{1}{2}\int\limits_{-T}^{0}dt\left[ \frac{||{\bf \dot{M}-(}\alpha
-1)\left( {\bf M}^{2}-{\bf \Pi }tr{\bf M}^{2}\right) ||^{2}}{C_{\eta
}\epsilon \rho ^{-2}}\right.  \label{20} \\
&&\left. +\frac{tr [ ({\bf \dot{\rho}%
-\rho M)}\left ( C_{\perp }^{-1}(1-{\hat \rho} {\hat \rho}^{\dagger }) +
C_{||}^{-1}{\hat \rho} {\hat \rho}^{\dagger }\right)  
({\bf \dot{\rho}-\rho M})^{\dagger} ] }{\rho
^{2}\sqrt{tr{\bf MM}^{\dagger }}}\right]  \nonumber
\end{eqnarray}
where $||X||^{2}\equiv tr{\bf XX}^{\dagger }$, ${\hat \rho }\equiv \rho /||\rho||$
and $C_{||}\rightarrow 0$, as assumed before. 

This path integral form invites a semi-classical approximation \cite
{SS94,Mig,MC} which estimates the integral via the saddle point $G_{T}({\bf %
M,\,\rho |M}^{\prime }{\bf ,\,\rho }^{\prime }){\bf \ \sim }exp[-S_{c}({\bf %
M,\,\rho |M}^{\prime }{\bf ,\,\rho }^{\prime })]$ given by the minimal
action $S_{c}$ along the ''classical'' trajectories connecting the
prescribed initial and final points (in time $T$) and obeying the
Euler-Lagrange variational equations. Moreover, for each final point $({\bf %
M,\,\rho })$ there exists a unique $S=0$ trajectory governed by the
deterministic part of Lagrangian dynamics (\ref{3a},\ref{3b}) which picks
out the Lagrangian preimage ${\bf M}^{\prime }={\tilde{{\bf M}}}({\bf M},%
{\bf \,\rho },-T),{\bf \,\rho }^{\prime }={\bf \tilde{\rho}}({\bf M},\,{\bf %
\rho },-T)$ as an initial condition. If the small scale generated stochastic
component of the dynamics were small $C_{\perp },\,C_{\eta }\rightarrow 0$
these deterministic Lagrangian trajectories would control the Green's
function. Since the probability is constant along the zero action trajectory
the PDF of the final ${\bf M,\rho }$ is determined by the probability of its
Lagrangian preimage ${\bf M}^{\prime }={\tilde{{\bf M}}}({\bf M},\,{\bf \rho 
},-T)$ at the integral scale where the PDF is assumed to be Gaussian. Crude
as this {\it zero action} approximation is, it is the natural zeroth order
calculation and will provide some physically interesting insights as we
shall see below. The full semi-classical analysis will be deferred to a
forthcoming publication.

\section{Probability Distribution Functions and Statistical Geometry.}
To make contact with the numerical results we shall use the "poor man's" zero action
approximation introduced in the previous section and 
according to which the probability of given $\bf M$ observed on a tetrad $\rho$ 
in the inertial range is equal to the probability of its integral scale preimage. 
The latter is found by integrating the deterministic part of the equations of motion 
(which generates zero action
trajectories) from the observation point {\it backward in time}:
\begin{mathletters}
\begin{equation}
{\frac{d}{dt}}{\bf M}\,=\,-\,(\alpha -1)({\bf M}^{2}-{\bf \Pi }tr{\bf M}^{2})
\label{22a}
\end{equation}
\begin{equation}
{\frac{d}{dt}}{\bf g}=-{\bf g}{\bf M}-{\bf M}^{\dagger }{\bf g}^{\dagger
}\,-\,\beta \sqrt{tr{{\bf M}{\bf M}^{\dagger }}}\,({\bf g}-{\frac{{\bf 1}}{3}}%
\,tr\,{\bf g})  \label{22b}
\end{equation}
where ${\bf g}\equiv {\bf \rho }^{\dagger }{\bf \rho }$. The $\beta $ term
has been added to reintroduce the isotropisation effect due to the
transverse small scale fluctuations already in the deterministic
approximation. This may be thought of as a Mean Field treatment of the $%
C_{\perp }$ term in (\ref{7}) which is physically more appropriate than the
formal $C_{\perp },C_{\eta }\rightarrow 0$ deterministic limit of (\ref{20}%
). Dimensionless constants $\alpha $ and $\beta $ will serve as model
parameters. Equations (\ref{22a},\ref{22b}) will be integrated back in time
until $tr{\bf g}=\rho ^{2}$ reaches the integral scale, yielding for the PDF:
\end{mathletters}
\begin{equation}
P({\bf M},\,{\bf g})\,{\bf \sim }\,exp\,\left[ -\,\frac{tr\left( {\bf M}%
_{L}^{\prime }({\bf M},\,{\bf g}){\bf M}_{L}^{\prime \dagger }({\bf M},\,%
{\bf g})\right) }{(\epsilon L^{-2})^{2/3}}\right]  \label{23}
\end{equation}
with ${\bf M}_{L}^{\prime }({\bf M},\,{\bf g})$ ( and the time of flight $T$%
) fixed implicitly by $tr\left[ {\tilde{{\bf g}}}({\bf M},\,{\bf g}%
,-T)\right] =L^{2}$. To the extent that the non-trivial PDF in this
approximation arises as a non-linear mapping of the initially Gaussian
variables, our construction here is reminiscent of Kraichnan's ''Mapping
Closure''\cite{RHK90,SJO90}.

Let us now construct the PDF of the $R,Q$ invariants for the isotropic
tetrad of radius $r$. It is convenient to consider the elliptic $D>0$ and
the hyperbolic $D<0$ regions separately and use different parameterizations
of the ${\bf M}$ matrix:

a) for $D>0$
\begin{mathletters}
\begin{equation}
{\bf M}=\left( 
\begin{array}{lll}
\lambda & \Delta e^{\gamma } & \omega _{2} \\ 
-\Delta e^{-\gamma } & \lambda & \omega _{1} \\ 
0 & 0 & -2\lambda
\end{array}
\right)  \label{24a}
\end{equation}
and b) for $D<0$
\begin{equation}
{\bf M}=\left( 
\begin{array}{lll}
\lambda +\Delta & \omega _{3} & \omega _{2} \\ 
0 & \lambda -\Delta & \omega _{1} \\ 
0 & 0 & -2\lambda
\end{array}
\right)  \label{24b}
\end{equation}
where $\omega _{a}$ refers to vorticity and in (\ref{24a}) $\omega
_{3}\equiv 2\Delta \cosh \gamma $. The invariants are a) $R=2\lambda
(\lambda ^{2}+\Delta ^{2})$ and $Q=\Delta ^{2}-3\lambda ^{2}$ and b) $%
R=2\lambda (\lambda ^{2}-\Delta ^{2})$ and $Q=-\Delta ^{2}-3\lambda ^{2}$
and the strain tensor is ${\bf s}\equiv ({\bf M}+{\bf M}^{\dagger })/2$. It
is straightforward to numerically integrate (\ref{22a},\ref{22b}) starting
with given ${\bf M}$ and ${\bf g}=r^{2}{\bf 1}$ in time until ${\frac{1}{3}}%
Tr{\bf g}=L^{2}$ at which point $P({\bf M},{\bf g})$ is assigned via (\ref
{23}). However, to determine $P_{r}(R,Q)$ one must integrate $P({\bf M})$
over a) $\gamma $, $\omega _{1,2}$ and b) over ${\vec{\omega}}$. (Note the
Jacobian: $\int d{\bf M}\delta (tr{\bf M})\,=\,\int dRdQd{\vec{\omega}}%
\,=\,\int d\lambda d\Delta d{\vec{\omega}}\sqrt{{|D|}}$ and $d\Delta d\omega
_{1}\,=\,2\Delta \gamma \sinh \left[ \gamma \right] d\gamma d\Delta $).

The task is simplified within the saddle approximation where the integration
is reduced to minimization \cite{ft9} of $logP({\bf M},%
{\bf r})$ with respect to the integration variables which we carry out
numerically via an ''amoeba'' algorithm \cite{NR}. Over the whole $R,Q$
plane we find that the saddle point is at a) $\gamma =\omega _{1,2}=0$ and
b) ${\vec{\omega}}=0$. In addition for the special case of $D=0$, when $%
\lambda $ is the only non-zero parameter in (\ref{24a},\ref{24b}) the
trajectory and the $P({\bf M},{\bf g})$ can be computed analytically (see
Appendix A). The resulting distribution (for different $r$) is displayed on
Fig. 5a,b. $P(Q,R)$ exhibits a long (but Gaussian ${\bf \sim }exp(-a\lambda
^{2}r^{2\gamma _{+}})$) ridge - the ''Vieillefosse tail'' - along the $D=0$
line in the $R>0,\,Q<0$ quadrant and a valley of low probability approaching
the origin from the $R>0$ side. This structure appears because the backward
in time trajectory of the point on the $D=0$ separatrix converges to the
origin and maps to a highly probable, small ${\bf M}^{\prime }$ integral
scale preimage, whereas the trajectories originating in the low probability
gulf in their time reversed dynamics are swept westward past the origin into
the improbably large ${\bf M}^{\prime }$ region. The PDF $P_{r}(R,Q)$
evolves continuously with decreasing $r/L$ away from the Gaussian (which
appears non-trivial in the $R,\,Q$ variables) shown for comparison on Fig.6.
The appearance of the PDF tail along $D=0,\,R>0$ and the the trend of $r/L$
dependence are reminiscent of those for the PDF obtained from the
Navier-Stokes. Yet, both high probability ridge and the low probability
valley found in the present deterministic approximation are strongly
exaggerated. There is a good reason to expect that the effect of the
fluctuations will be strong in these regions: the narrow ridge should be
largely ''washed out'' as the asymptotic behavior of the tails and low
probability regions is clearly dominated by fluctuations. Yet again, the
rather complex $R,Q$ dependence of the PDF and the crude qualitative
similarity of the model PDF with the results of the DNS merits a detailed
discussion of the underlying kinematics and dynamics.

Let us compute the distribution in $R,Q$ plane of the average enstrophy $
\omega ^{2}$ and enstrophy production ${\vec{\omega}}\cdot {\bf s}\cdot {%
\vec{\omega}}={\frac{1}{2}}Tr({\bf M}+{\bf M}^{\dagger })({\bf M}-{\bf M}%
^{\dagger })^{2}$ which measures the rate of vortex stretching\cite{MY}. The
enstrophy density is defined by $e(R,Q)\,=\,\int d{\vec{\omega}}\omega
^{2}P_{r}(R,Q,\omega )$ but in order to save computer time will only be
evaluated in the saddle approximation by varying the integrand w.r.t. ${\vec{%
\omega}}$. For all of the hyperbolic region the saddle of the integral
occurs at a non-zero values of $\gamma ,\,\vec{\omega}$ parameters (\ref{22a}%
,\ref{22b}). The result is presented in Fig.7. We see that the average
enstrophy peaks at $R=0$ and small positive $Q$. This is explained by noting that
the conditional average $\left\langle {\vec{\omega}}^{2}|R,Q\right\rangle $
grows like $Q$ (at least for $R\approx 0$ and $Q>0$) because $Q\,=\,({\frac{1%
}{2}}{\vec{\omega}^{2}}-Tr\,{\bf s}^{2})/2$, while $P(R,Q)$ falls off. The
average enstrophy production $\sigma (R,Q)\equiv \int d{\vec{\omega}}\,{\vec{%
\omega}}\cdot {\bf s}\cdot {\vec{\omega}}\,P_{r}(R,Q,\vec{\omega})$ is also
dominated by a non-trivial saddle and has the $R,Q$ dependence shown in
Fig.8. We observe that enstrophy is produced predominantly in the upper left
quadrant of the $R,Q$ plane, it is (partially) destroyed in the upper right
quadrant and there is weak vortex stretching in the $D=0$ tail. The tail
region is not well resolved on Fig.8 and the positive enstrophy production
is confined to the narrow strip delineated by the zero contours. This strip
becomes progressively more and more narrow as one looks at smaller scales, $%
r $, (which is why we limited our figures to rather large $r/L,$) however, the
existence of positive vortex stretching in the tail domain was verified
analytically via the arguments presented later in this section (and in
Appendix A). Next we compute the
average strain skewness, $S_{3}\equiv \int d{\vec{\omega}}\,tr\,{\bf s}%
^{3}P_{r}(R,Q)$, which is the object associated with energy transfer \cite
{MY}, and find that it is strongly localized in the $D=0$ Vieillefosse tail
as shown in Fig.9. (Curiously, we find that in the elliptic region, $Q>0$, the strain skewness
changes sign three times - a fact that can be understood on the basis of the
structure of the ${\bf M}$ tensor given by (25).) Note that the energy transfer term in (\ref{9}) is
actually $Tr\,{\bf M}^{2}{\bf M}^{\dagger }\,=\,Tr{\bf s}^{3}-{\frac{1}{4}}{%
\vec{\omega}}\cdot {\bf s}\cdot {\vec{\omega}}$ so that vortex stretching
also contributes to the energy flux.\cite{ft10} Vortex stretching however does not dominate the energy
transfer: while most of the enstrophy production is in the upper left
quadrangle, Fig.8, the energy flux distribution $F(R,Q)\,\equiv -\alpha \int
d{\vec{\omega}}\,Tr\,{\bf M}^{2}{\bf M}^{\dagger }P_{r}(R,Q,{\vec{\omega}})$
is localized in the $D=0$ tail where the negative $Tr\,{\bf s}^{3}$ lives,
see Fig.10. Thus we arrive at the conclusion that from the energy transfer
point of view, the active regions are dominated by the strain and not
vorticity \cite{JIM93,Tsi97}. Also of interest is the appearance of distinct
regions of weak {\it negative} energy transfer: e.g. the $R,Q>0$ quadrangle
where both $\sigma $ and the energy flux are negative\cite{RE}.

Remarkably, as seen from the comparison of Fig.7-10 and
Figs. 11-14 the DNS exhibits a rather similar distribution of average enstrophy, 
enstrophy production and strain skewness in the $Q,R$ plane. Both DNS and the model
have positive enstrophy production in the upper left quadrant and in the $D=0$, $R>0$ tail, and negative
$\sigma$ in the upper right quadrangle; both have strain skewness strongly confined to the $D=0$, 
$R>0$ tail, and both exhibit the dominance of the strain skewness in the energy transfer. Furthermore,
there is a clear correspondence of the positive and negative regions of $tr {\bf s}^3$ (Fig.9 and 13)
and energy flux (Fig.10 and 14).

Much of this behavior can be understood by considering
the ''statistical geometry'' of the flow (see Ashurst et al\cite{AKKG},
Constantin \cite{Cons} and Tsinober et al\cite{Tsi97} for excellent
discussions) and is inherited from the RE dynamics. Most of the vortex
stretching occurs for $R<0,\,Q>0$ where the vorticity is aligned with the
large positive eigenvector of the strain; this is immediately evident from
the structure of the ${\bf M}$ matrix at the saddle point controlling $%
\sigma (R,Q)$: 
\end{mathletters}
\begin{equation}
{\bf M}_{\sigma }=\left( 
\begin{array}{lll}
\lambda & \Delta & 0 \\ 
-\Delta & \lambda & 0 \\ 
0 & 0 & -2\lambda
\end{array}
\right)  \label{25}
\end{equation}
where\cite{ft11}$ \lambda <0$, since the off-diagonal
elements of (24) define the vorticity along the direction with the strain eigenvalue $-2\lambda >0$. The
tetrad dynamics then takes the ${\bf M}$ field into the $R\approx 0,\,Q>0$
region where $\lambda $ and hence the stretching rate $\sigma $ vanish
before changing sign. The magnitude of vorticity reaches a maximum. In the
peak vorticity region ${\bf M}$ tensor has the form: 
\begin{equation}
{\bf M}_{e}=\left( 
\begin{array}{lll}
\lambda & \Delta e^{\gamma } & 0 \\ 
-\Delta e^{-\gamma } & \lambda & 0 \\ 
0 & 0 & -2\lambda
\end{array}
\right)  \label{26}
\end{equation}
with $\lambda \ll \Delta $ and $\gamma \ll 1$, so that the flow is
essentially two dimensional with vorticity aligned with the nearly neutral
strain direction.

The
alignment of intense vorticity with the intermediate strain axis is well
known \cite{EDS81,Kerr,AKKG,JIM93,Tsi97} and its reappearance in the model
is encouraging. The fate of the typical $2D$ vorticity is to implode under
the action of the contracting strain (brought about by the low pressure of
the vortex) \cite{ft12}.
The energy is transfered to larger scale. 

To conclude we examine the geometry of ${\bf M}$ in the $D=0$ tail. Here the
predominant ${\bf M}$ configuration is 
\begin{equation}
M_{s}\,=\,\left( 
\begin{array}{lll}
\lambda & 0 & \omega \\ 
0 & \lambda & 0 \\ 
0 & 0 & -2\lambda
\end{array}
\right)  \label{27}
\end{equation}
which leads to the strain eigenvalues 
\begin{equation}
s_{\alpha }\,=\,\lambda ;\;\;\,\,s_{\beta }\,=\,{\frac{3}{2}}\,\lambda \,%
\sqrt{{{1+{\frac{\omega ^{2}}{9\lambda ^{2}}}}}}{-{\frac{\lambda }{2}}}
\label{28}
\end{equation}
and $s_{\gamma }=\,-s_{\alpha }-s_{\beta }$. The vorticity is aligned with
the intermediate strain axis, which is stretching since the $s_{\beta }>0$
consistent with the negative strain skewness $Tr\,{\bf s}^{3}\,=\,3s_{\alpha
}s_{\beta }s_{\gamma }$ The normalized $\beta $-strain \cite{AKKG,BO98} $%
\left\langle {\frac{s_{\beta }}{|{\bf s}|}}|\lambda \right\rangle $ as a
function of $\lambda $ along the $D=0$ line increases as $\left\langle {%
\frac{\omega ^{2}}{\lambda ^{2}}}|\lambda \right\rangle $ goes down with
increasing $\lambda $ as it is shown below in the Section and confirmed by
Fig.15a,b for the DNS. To the extent that $P_{r}(R,Q)$ is peaked at $R=Q=0$
and hence at $\lambda =0$ the most probable configurations have $s_{1}\gg
s_{2}\approx 0$ and are close to plane shear or $2D$ vorticity. Note that
these most probable configurations have $Tr\,{\bf M}^{2}\,\approx 0$ or $Tr%
{\bf s^{2}}\,\approx {\frac{1}{2}}{\vec{\omega}}^{2}$ which means
approximate $local$ homogeneity - a much stronger statement than the
homogeneity on average, i.e. $\left\langle Tr\,{\bf M}^{2}\right\rangle \,=0$%
! This must have serious implications for the pressure. 

The exagerated non-Gaussianity (i.e. Vieillefosse tail) of the deterministic model aside,
there are also other qualitative differences between the results of the model
and the DNS: e.g. the vorticity
density in the DNS peaks at the origin and is skewed in the upper half
plane, towards $R<0$. This effect is even stronger for
the vorticity distribution measured in the dissipative range: 
an asymmetry also visible in $P(R,Q)$ shown on
Fig.3a. 
This can be understood as a result of enstrophy dissipation by
viscosity which shifts the locus of null enstrophy growth and hence maximal enstrophy
to configurations where of vorticity is still aligned with positive strain. 


Let us further explore the results of the deterministic approximation. The
exact solution on the $D=0$ line (see Appendix A) exhibits strong asymmetry:
even though the PDF is Gaussian along the $D=0$ line both for $R>0$ and $R<0$%
, $ln\,P(\lambda )\sim -\lambda ^{2}(r/L)^{4(1-\alpha )h_{\pm }(\beta )}$
the characteristic exponents on the two sides are different, $h_{+}(\beta
)>h_{-}(\beta )$. It is the $h_{-}$ exponent which controls the behavior of
low moments including the 3d and in order to impose constant energy flux we
must require $(1-\alpha )h_{-}(\beta )=2/3$ which forces K41 scaling on the
''head'' of the PDF in Fig.5a,b. The constant flux condition thus fixes a
specific relation (determined in the Appendix A) between the model
parameters: as $\beta $ increases from 0 to $\approx .3$, $\alpha $
decreases from $1/3$ to 0.

However, while the scaling in the bulk of the PDF is K41, the scaling in the
Vieillefosse tail is anomalous. Introducing rescaled variable ${\tilde{%
\lambda}}=\lambda r^{{\frac{2}{3}}}$, $\;\;{\tilde{\Delta}}=\Delta r^{{\frac{%
2}{3}}}$, and ${\tilde{\omega}}=\omega r^{{\frac{2}{3}}}$ we find that the ${%
\tilde{\Delta}}$ dependence of the action in the hyperbolic vicinity of the
tail (which determines its width in the $R,Q$ plane) has the asymptotic form: 
\begin{equation}
S_{+}({\tilde{\lambda}},{\tilde{\Delta}},\omega =0)\,\sim {\tilde{\lambda}}%
^{2}r^{2\gamma _{+}}\,+\,{\tilde{\Delta}}^{2}r^{-2\eta }f_{\Delta }\left( {%
\frac{{\tilde{\Delta}}}{{\tilde{\lambda}}}}r^{-\eta -\gamma _{+}}\right)
\label{29}
\end{equation}
where the scaling function $f_{\Delta }(0)=const$ and $f_{\Delta }(x){\bf %
\sim }x^{\zeta }$ for $x\gg 1$ with $\eta =\gamma _{+}\zeta /(2-\zeta )$.
Similarly, the vorticity distribution on the $D=0$ ($R>0$) line is governed
asymptotically by 
\begin{equation}
S_{+}({\tilde{\lambda}},{\tilde{\Delta}}=0,{\tilde{\omega}})\,\sim \,{\tilde{%
\lambda}}^{2}r^{2\gamma _{+}}\,+\,{\tilde{\omega}}^{2}r^{-2\eta ^{\prime
}}f_{\omega }\left( {\frac{{\tilde{\omega}}}{{\tilde{\lambda}}}}r^{-\eta
^{\prime }-\gamma _{+}}\right)  \label{30}
\end{equation}
where $f_{\omega }(x)\sim x^{\delta }$ for $x\ll 1$ and $f_{\omega }(x)\sim
x^{\delta -3}$ for $x\gg 1$ with $\eta ^{\prime }=\gamma _{+}(3-\delta
)/(\delta -1)$. The anomalous exponents depend on $\beta $: $\gamma
_{+}\,\equiv 2(1-\alpha )(h_{+}(\beta )-h_{-}(\beta ))$ varies from 0 at $%
\beta =0$, ($\alpha =1/3$) to 1/2 at $\beta \approx .3$, $\alpha =0$. Over
the same range $\eta $ varies from 1 to 3/2. 
The exponents as a
function of $\beta $ are tabulated in the Appendix A

It follows that the conditional enstrophy in the Vieillefosse tail behaves
for large $\lambda $ as: 
\begin{equation}
<{\tilde{\omega}}^{2}|{\tilde{\lambda}}>\sim \,({\tilde{\lambda}})^{{\frac{%
2\delta }{2+\delta }}}\;r^{{\frac{{2}\gamma {_{+}\delta }}{{2+\delta }}}%
+2\eta ^{\prime }}  \label{31}
\end{equation}
implying an interesting non-trivial scaling relation between the strain and
vorticity of the velocity gradient sheets associated with the high energy
transfer regions.

We can also compute the contribution of the $D=0$, $R>0$ tail to the monts
of the velocity gradient. E.g. we compute 
\begin{equation}
\left\langle {\tilde{\lambda}}^{n}\right\rangle _{tail}\,=\,N\int d{\tilde{%
\lambda}}^{3}\int d{\tilde{\Delta}}^{2}\int d{\tilde{\omega}}{\tilde{\lambda}%
}^{n}\,e^{-S_{+}({\tilde{\lambda}}r^{\gamma _{+}},{\tilde{\Delta}}r^{-\eta
},{\tilde{\omega}}r^{-\eta ^{\prime }})}\,\sim r^{2\eta +3\eta ^{\prime
}-(n+3)\gamma _{+}}  \label{32}
\end{equation}
where the normalization factor $N\sim 1$ because it is dominated by the
''head'' of the PDF with K41 scaling, already factorized explicitly. Thus we
conclude that the tail contribution is important only for sufficiently large 
$n$, when $n>n_{c}=(2\eta +3\eta ^{\prime })/{\bf \ }\gamma _{+}-3$ and the
anomalous scaling becomes dominant over the normal contribution of the bulk
of the PDF. We find that in the limit of $\beta \rightarrow 0$ and $\alpha
\rightarrow 1/3$, $\;n_{c}\rightarrow \infty $ and the shape of the PDF,
although non-Gaussian, becomes independent of the scale. This limit recovers
the Kolmogorov 41 theory. The intermittency effect is maximized as $\beta
\rightarrow .3$ and $\alpha \rightarrow 0$, where anomalous scaling appears
for $n>6$.

It is clear that although the PDF found in the present deterministic
approximation of the model is on the whole far from Gaussian (imposed on
integral scale) the approximation underestimates the intermittency effects.
E.g. it predicts no deviations from K41 in the vorticity dominated $Q>0$
region, in contrast to the DNS result which in that region exhibits $r$%
-dependence associated with the development of an exponential (or
sub-exponential) tail on smaller scales. It is equally clear that the
description of such a tail is beyond the currently employed approximation.
After all, the asymptotic behavior of the PDF, i.e. the statistics of large
fluctuations, is dominated by fluctuations. One does however expect to find
the exponential asymptotics \cite{SS94} once the fluctuations are accounted
for via a proper semi-classical calculation. There thus appears to be two
mechanisms contributing to the intermittency: 1) the deterministic
non-linear interactions in the energy transfer region as indicated by our
present calculation and 2) the effect of small scale fluctuations
responsible for the exponential asymptotics of the PDF. The latter mechanism
is analogous to the one responsible for the intermittency of the Passive
Scalar\cite{RHK94,SS95,WG4,GK}.

\section{Conclusions.}
In the preceding Sections we have introduced and began to analyze the
phenomenological model of inertial scale velocity fluctuations defined
through the velocity gradient tensor coarse-grained over a region specified
by a tetrad of points. The dynamics of this field was decomposed into the
non-linear deterministic component representing local same-scale
interactions and a Gaussian stochastic component with embedded K41 scaling
representing interactions non-local in space and the incoherent contribution
of the velocity fluctuations from scales smaller than that of the tetrad.
The deterministic component is closely related to the RE dynamics 
of Vieillefosse \cite{VFS} model. The latter, although marred by the
unphysical finite time singularity, \cite{ft13}
has been an appealing candidate description for the dynamics of the local
velocity gradient \cite{AKKG,SJO91,BO98}. A novel aspect of our model is the
elevation of the velocity gradient dynamics to the coarse-grained level and
the introduction of the tetrad tensor (${\bf \rho }$) dynamics which
explicitly introduces the scale and the measure of anisotropy generated
through the strain induced distortion of the Lagrangian volume. This allows
to construct an anisotropic model of the coarse grained pressure Hessian
which eliminates the finite time singularity from the deterministic dynamics
by suppressing the work done by the pressure on the distorted fluid element.
Furthermore, the explicit appearance of the current scale allows to build in
K41 spectrum in the stochastic component \cite{ft14} of the dynamics. 
An important reason for working with the coarse grained
field is that in the dynamics of the purely inertial range fields the direct
contribution of the viscosity can be neglected \cite{ft15}. Instead, the ultimately
viscous dissipation is incorporated through the effect of the incoherent
small scales acting through the eddy damping $D_{nl}+D_{\perp }$ terms in
the model. 
The deterministic dynamics plays the key role in transferring energy
down scale. This transfer occurs due to the volume preserving distortion of
the fluid element which leads to the reduction of at least one of its
principle axis.

Our heuristic derivation of the model was fortified by the numerical test of
(\ref{3a}) through the construction of the conditional flow for the
invariants of ${\bf M}$. The deterministic part of (3a) appears to be quite
close to the conditional flow at least for isotropic tetrads ${\bf \rho }=%
{\bf 1}\,r$. It will be important to extend the numerical study to
anisotropic $\rho $ and to find a way of examining the stochastic
contribution to the dynamics. The validity of the local approximation of the
pressure Hessian \cite{SJO91} and the neglect of non-local correlations
remain the key issues.

The tetrad model respects K41 scaling (${\bf M}\rightarrow b{\bf M}$, $\;%
{\bf \rho }\rightarrow b^{-3/2}{\bf \rho }$), both on the operator (\ref{7})
and the integral scale boundary condition (\ref{16}) levels. Yet, as
illustrated by our crude deterministic approximation the resulting PDF has
non-trivial, model parameter dependent anomalous scaling. The Kolmogorov
scaling has to be re-imposed on the level of the 3d moment by choosing the
parameters so as to make the energy flux scale independent. Except for one
particular limit ($\beta \rightarrow 0$, $\alpha \rightarrow 1/3$) where the
PDF has K41 behavior, high moments exhibit anomalous scaling. Remarkably,
while this anomalous scaling has little to do with the Kolmogorov-Obukhov
arguments \cite{MY,UF} it does originate in the domain of energy transfer:
the Vieillefosse tail.

The deterministic approximation employed in Section 5 grossly overestimates
the extent of the Vieillefosse tail in $P(Q,R)$ but it does bear resemblance
to the PDF observed in the DNS. It also generates plausible distributions of
enstrophy, enstrophy production and strain skewness. The present analysis
provides a clear dissection of the high enstrophy and the high enstrophy
production regions, identifies the difference in the vorticity- strain
alignment in the two regions \cite{Tsi97} and exhibits their dynamical
connection. As explained in Section 3, much of this sensible phenomenology
was inherited from the RE dynamics \cite{AKKG}. 
The elimination of the Vieillefosse finite time singularity was however
essential in order to have a model with stationary statistics. 

The energy transfer for isotropic $\rho $ configurations occurs via
non-linear eddy damping \cite{Bard,BO98} $tr{\bf M}^{2}{\bf M}^{\dagger }$
term (\ref{9}), which combines contributions of strain skewness and vortex
stretching. Within our present crude deterministic approximation, the energy
transfer occurs largely in the Viellefosse tail and is due to large negative 
$tr{\bf s}^{3}$. In this region strain has two positive eigenvalues and
Lagrangian volumes are deformed into pancakes or ribbons, i.e. this is the
region of sheet formation \cite{Betch}. In contrast, the vortex filaments
are generated in the $Q>0,\;R<0$ quadrant where enstrophy production is
peaked. This region does not contribute as much to the energy flux.
Furthermore, the maximal vorticity region is characterized by nearly $2D$
configurations and does not contribute at all - a notion consistent with
recent numerical results \cite{JIM93}. Yet, whereas we are optimistic about
correct description of the high $-tr{\bf s}^{3}$ tail in the model, the
correct description of the high $\sigma $ region may be more difficult
because of the importance of long-range strain in vortex stretching.

In order to enable a more quantitative comparison of the model and the DNS
the treatment of the model must include the effect of the fluctuations which
control the asymptotic behavior of the PDF. Our present, deterministic,
approximation overestimates the PDF in the narrow Vieillefosse tail, while
underestimating the asymptotic behavior of the PDF elsewhere in the $Q,R$
plane. Both effects are due to the neglect of the fluctuations. Since the
relevant Fokker-Planck equation lives in the $d^{2}-1+d(d+1)/2=14$
dimensional ${\bf M}$,{\bf \ }$\rho $ space the direct numerical approach
seems out of question. However one may hope to make progress with the
semi-classical analysis of the path integral representation (\ref{20}) along
the lines of \cite{SS94,MC}. This approach is valid for the calculation of
the PDF tails and associated anomalous scaling. We expect that the
fluctuation effects will change the Gaussian decay of the PDF tails found in
the deterministic approximation to the exponential,\cite{ft16}
e.g.: $\lambda
^{-a}\,e^{-c\lambda r^{b}}$.

An interesting simplification appears in the limit of strong transverse
diffusion $C_{\perp }>>1$ which corresponds to the physically sensible
regime of strong re-isotropisation of the tetrads arising from the action of
incoherent small scale fluctuations. In that limit the PDF becomes nearly
uniform over the $\rho ^{2}=const$ shells and can be projected onto the $s$
and $d$ representations of $SO(9)$ acting on $\rho _{i}^{a}$:
\begin{equation}
P({\bf M},{\bf \rho })\,\approx \Psi ({\bf M},\rho ^{2})\,+\,{\frac{1}{{%
2d^{2}\,C_{\perp }}}}\;(\rho _{i}^{a}\rho _{j}^{b}-{\frac{1}{d^{2}}}\rho
^{2}\delta ^{ab}\,\delta _{ij}\,)\;\Phi _{ij}^{ab}({\bf M},\rho ^{2}).
\label{33}
\end{equation}

Curiously, this nearly isotropic approximation doubles as a $d>>1$ expansion 
\cite{FFR,WG4,Yakh} because the $d$-wave mode of $SO(d^{2})$ is suppressed
by the $l(l+d^{2}-2)$ total angular momentum factor with $l=2$. This
expansion allows a considerable simplification of (\ref{7}) and (\ref{20}).

Several related models are worth mentioning. The present model appears to be
the real space, Lagrangian counterpart of the momentum space, Eulerian
''shell'' model proposed by Siggia \cite{EDS77}. 
Yet, the geometrical and statistical implications of that model have not
been fully explored and we do not at present understand the relative merits
of the assumptions involved in the two models. The logic which led to (\ref
{3a},\ref{3b}) can and has been applied to the Passive scalar problem, in
which case one would replace ${\bf M}$ by a Gaussian random field. This
would lead to a PS model of the type considered in \cite{SS95,SS96} closely
related to the Kraichnan's \cite{RHK68} model. It would be interesting also
to explore whether the 2D version of (\ref{3a},\ref{3b}) could provide a
sensible description of 2D turbulence where the physics is very different.
In that case the deterministic left hand side of (\ref{3a}) would vanish for
isotropic ''triads''. We do not presently know if the model can generate the
inverse cascade. Finally, if the tetrad model provides a
reasonable description of turbulent fluctuations in the homogeneous case, it
will be interesting to attempt to generalize it to the inhomogeneous and
anisotropic case: e.g. one could study the statistics of tetrads moving away
from the wall in the boundary layer.

Last but not least, two remarks concerning the relation with the experiment.
The traditional approach to turbulence and most of the existing data
involves velocity difference at two points. This two point statistics can be
extracted from the tetrad statistics by averaging over ''unobserved''
variables $P({\vec{v}},{\vec{\rho}}_{1})=\int d{\vec{\rho}_{2}}\int d{\vec{%
\rho}_{3}}\int d{\bf M}\delta (\rho _{1}^{a}M_{ab}-v_{b})P({\bf M},\rho )$.
Conversely, it would be very interesting to study tetrad statistics
experimentally. For that we need a velocity measurement at 4 points in the
inertial range, which can be obtained from 3 (crossed wire) probes (say a
fixed probe at the origin and two movable probes at $(x,y,z)=(0,\sqrt{{3}/2}%
r,\pm 1/2r)$ in a flow with $<v_{x}>\neq 0$. The time lag (on the static
probe) then can be used to provide the 4th measurement \cite{MPSSW,MW}. The
PDF coarse grained velocity gradient may then be obtained. Quite
independently of the predictions of the present model, the $(R,Q)$-plane
density of the 2nd and 3d order invariants (e.g. Figs. 8-14) are quite
illuminating in dissecting the role of vorticity and strain in intermittency
and the possible difference in their anomalous scaling.

We conclude finally that despite its relative simplicity the tetrad model is
surprisingly rich in physics, offering an insight both into the geometry and
dynamics as well as the statistical and scaling properties of the inertial
range fields. Many non-trivial statistical objects can be calculated in
terms of the 3 parameters of the present model. Further work and detailed
comparison with the DNS and the experiment will establish the degree of
success and failure of this model.

\section{Acknowledgements}
It is a pleasure to thank A. Libchaber,
E.D. Siggia and A. Tsinober for stimulating discussions. MC acknowledges support of
the R.H. Dicke fellowship and the hospitality of Bell Labs. We also greatfully
acknowledge the grant of computer time from IDRIS (France).

\appendix

\section{ Scaling in the deterministic limit}
Consider Lagrangian dynamics of $\hat{M}$ and $\hat{g}$ described by (\ref
{22a},\ref{22b}) for the following class of matrices forming a subspace in
the Elliptic region ($D\geq 0$) 
\begin{equation}
\hat{M}=\left( 
\begin{array}{lll}
\lambda & \Delta & 0 \\ 
-\Delta & \lambda & 0 \\ 
0 & 0 & -2\lambda
\end{array}
\right) ,\hspace{0.2in}\hspace{0.2in}\hat{g}=\left( 
\begin{array}{lll}
x & 0 & 0 \\ 
0 & x & 0 \\ 
0 & 0 & y
\end{array}
\right) .  \label{diagMP}
\end{equation}
We arrive at the system of equations 
\begin{eqnarray}
\frac{d\ln \left[ \lambda \right] }{\lambda dt} &=&\left( 1-\alpha \right) 
\frac{z-4-\mu ^{2}z}{z+2},  \label{meq} \\
\frac{d\ln \left[ \mu \right] }{\lambda dt} &=&\left( 1-\alpha \right) \frac{%
z+8+\mu ^{2}}{z+2},  \label{mueq} \\
\frac{dz}{\lambda dt} &=&-6z+sgn[\lambda ]\frac{\beta }{3}\sqrt{6+2\mu ^{2}}%
(1+2z)(1-z),  \label{zeq} \\
\frac{d\ln \left[ G\right] }{\lambda dt} &=&\frac{4(1-z)}{1+2z},  \label{Geq}
\end{eqnarray}
with $\mu \equiv \Delta /\lambda $, $z\equiv x/y$ ($z(0)=1$), 
$G\equiv tr[\hat{g}]=2x+y$. 
This system is integrable in quadratures if either $\beta =0$ or $\Delta
(0)=0$ (and therefore, $\Delta (t)=0$, for any other $t$). We shall now
calculate the effective action $S\equiv -\ln \left[ P({\bf M,g})\right] $%
(see (\ref{23})) for these two cases.

\subsection{Elliptic region, $\beta =0$.}

Integration of the system (\ref{meq}-\ref{Geq}) gives 
\begin{eqnarray} & &
\lambda ^{\prime }{}^{2} \!=\!\left[ \frac{3x^{\prime 2}r^{2}}{x^{\prime
3}\!+\!2r^{6}}\right] ^{1-\alpha } \label{A6}\\  &&\times
\left( \lambda ^{2}\!-\!\Delta ^{2}3^{\alpha -1}\left[ \left( \frac{%
r^{2}}{x^{\prime }}\right) ^{1-\alpha }\!\!\!\hspace{0.01in}_{2}F_{1}\!\left( \frac{%
1\!-\!\alpha }{3},\alpha ,\frac{4\!-\!\alpha }{3},\frac{-2r^{6}}{x^{\prime 3}}%
\right)\!-\hspace{0.01in}_{2}F_{1}\!\left( \frac{1\!-\!\alpha }{3},\alpha ,
\frac{4\!-\!\alpha }{3},-2\right) \right] \right) , \nonumber\\ &&
\Delta ^{\prime } =\Delta \left( \frac{r^{2}}{x^{\prime }}\right)
^{1-\alpha },\hspace{0.2in}y^{\prime }=\frac{r^{6}}{x^{\prime 2 }},
\label{ytxt}
\end{eqnarray}
where the '$^{\prime }$' notation stands to mark the $(-T)$ preimage. These
expressions allow to rewrite the effective action in terms of $r,\lambda $
and $\Delta $, providing $T$ is fixed by $tr[\hat{\rho}(-T)\hat{\rho}%
(-T)^{+}]=2x^{\prime }+r^{6}/x^{\prime 2}=3L^{2}=3$. There are two solutions
for $x^{\prime }$ realized separately depending on the sign of $\lambda
^{\prime }$.

First, consider the region with $\lambda $ being positive during the all
Lagrangian evolution. At $r\ll 1$, the respective value of the effective
action is 
\begin{equation}
\lambda ^{\prime }>0,\hspace{0.2in}S\rightarrow 3\lambda ^{2}\left( \frac{%
r^{2}}{2}\right) ^{1-\alpha }+\Delta ^{2}\left( \frac{\sqrt{3}}{r}\right)
^{2(1-\alpha )}.  \label{S+}
\end{equation}
We find that the action becomes infinite at $r\rightarrow 0$ in the domain.

If $\lambda (t)$ is negative (at least at the final stage of the backward in
time evolution) and $r\ll 1$ the effective action is 
\begin{equation}
\lambda ^{\prime }<0,\hspace{0.2in}S\rightarrow 3\left[ \lambda ^{2}+\Delta
^{2}3^{\alpha -1}\hspace{0.05in}_{2}F_{1}\left( \frac{1-\alpha }{3},\alpha ,%
\frac{4-\alpha }{3},-2\right) \right] \left( \frac{3r^{2}}{2}\right)
^{1-\alpha }.  \label{S-cl}
\end{equation}
The transition region between (\ref{S+}) and (\ref{S-cl}) shrinks with $%
r\rightarrow 0$. The crossover occurs at the intersection of the $\lambda
=\mu \Delta $ line in the $\lambda ,\Delta $ ( $Q,R$) plane. $\mu $, depends
on both $\alpha $ and $r$ and has the following asymptotic form: 
\begin{equation}
\mu\! \left|_{r\rightarrow 0}\right. \!\rightarrow\! \left\{ 
\begin{array}{c}
3^{\alpha -1}\left[ 2^{[\alpha -1]/3}\!\Gamma \left( \frac{4-\alpha }{3}%
\right) \!\Gamma \left( \frac{4\alpha -1}{3}\right) /[-3\Gamma (\alpha
\!-\!1)]-_{2}F_{1}\left( \frac{1-\alpha }{3},\alpha ,\frac{4\!-\!\alpha }{3}%
,-2\right) \right] ,\ \ \ \alpha >1/4; \\ 
3^{1/2-\alpha }(1\!-\!\alpha )r^{4\alpha -1}/[2^{\alpha +1}(1-4\alpha )(2\alpha
+1)],\ \ \ \alpha <1/4.
\end{array}
\right.  \label{beta}
\end{equation}
Therefore, a sector in the right part of $Q,R$ plane bounded by the $D=0$
line from below and the $\lambda =\mu \Delta $ one from above forms a low
probability gulf of the PDF.

\subsection{Zero discriminant line: $D=0$.}

For the $\delta =D=0$ line, the integration of (\ref{meq}-\ref{Geq}) yields 
\begin{eqnarray}
\ln \left[ \frac{G^{\prime }}{3r^{2}}\right] &=&4\int\limits_{z^{\prime
}}^{1}\frac{(1-\tilde{z})d\tilde{z}}{\left[ 6\tilde{z}-\beta \left( 1+2%
\tilde{z}\right) \left( 1-\tilde{z}\right) \sqrt{2/3}sgn\left[ \lambda
\right] \right] (1+2\tilde{z})},  \nonumber \\
\ln \left[ \frac{\lambda }{\lambda ^{\prime }}\right] &=&(1-\alpha
)\int\limits_{z^{\prime }}^{1}\frac{\left( 4-\tilde{z}\right) d\tilde{z}}{%
\left( 6\tilde{z}-\beta \left( 1+2\tilde{z}\right) \left( 1-\tilde{z}\right) 
\sqrt{2/3}sgn\left[ \lambda \right] \right) \left( 2+\tilde{z}\right) }.
\label{laint}
\end{eqnarray}
The dynamics (on $D=0$ line) does not change the sign of $\lambda $ so that $%
sgn[\lambda ]$ is constant.

For positive $\lambda $, $\lambda (t)$ is monotonically decreasing while $%
G(t)$ grows and $z(t)$ decreases. For small enough initial $G$($=3r^{2}\ll 1$%
), $z$ approaches $z_{+}$, 
\begin{equation}
z_{+}\equiv \frac{\beta -3\sqrt{6}+\sqrt{54-6\beta \sqrt{6}+9\beta ^{2}}}{%
4\beta }.  \label{z*}
\end{equation}
For $z^{\prime }$ close to $z_{+}$ one finds 
\begin{equation}
\frac{\lambda ^{\prime }}{\lambda }\rightarrow const*\left[ z^{\prime
}-z_{+}\right] ^{C_{1}^{+}(1-\alpha )},\hspace{0.2in}G^{\prime }\rightarrow 
\frac{const*r^{2}}{\left[ z^{\prime }-z_{+}\right] ^{C_{2}^{+}}},
\label{laGpos}
\end{equation}
where 
\begin{eqnarray}
C_{1}^{+} &=&\sqrt{3/2}\frac{4-z_+}{(2+z_+)\sqrt{54-6\beta \sqrt{6}+9\beta ^{2}}}
, \label{C1+}\\
C_{2}^{+} &=&\sqrt{24}\frac{1-z_+}{(1+2z_+)\sqrt{54-6\beta \sqrt{6}+9\beta ^{2}}}
.\label{C2+}
\end{eqnarray}
We get the following asymptotic behavior for the effective action for $D=0$
and $R,\lambda >0$: 
\begin{equation}
S\sim \lambda _{L}^{2}\sim \lambda ^{2}r^{4(1-\alpha )h_{+}},
\label{SZDLpos}
\end{equation}
with $h_{+}\equiv C_{1}^{+}/C_{2}^{+}$.

If $\lambda $ is negative, $\lambda (t)$ increases in absolute value while $%
z(t)$ decreases at the initial stage of backward in time evolution. The
dynamics changes ( $r$ is supposed to be very small) once $z$ crosses $4$
and $\lambda (t)$ starts moving towards the origin. $z$ keeps growing to
approach $z_{-}$, 
\begin{equation}
z_{-}=\frac{3\sqrt{6}+\beta +\sqrt{54+6\sqrt{6}\beta +9\beta ^{2}}}{4\beta }.
\label{zc2}
\end{equation}
For $z$ close to $z_{-}$ one finds 
\begin{equation}
\left| \frac{\lambda ^{\prime }}{\lambda }\right| \rightarrow const*\left[
z^{\prime }-z_{-}\right] ^{C_{1}^{-}(1-\alpha )},\hspace{0.2in}G^{\prime
}\rightarrow \frac{const*r^{2}}{\left[ z^{\prime }-z_{-}\right] ^{C_{2}^{-}}}%
,\hspace{0.2in}  \label{laGneg}
\end{equation}
where 
\begin{eqnarray}
C_{1}^{-} &=&\sqrt{3/2}\frac{z_--4}{(2+z_-)\sqrt{54+6\sqrt{6}\beta +9\beta ^{2}}}
,  \label{C1-} \\
C_{2}^{-} &=&\sqrt{24}\frac{z_--1}{(1+2z_-)\sqrt{54+6\sqrt{6}\beta +9\beta ^{2}}}
.  \label{C2-}
\end{eqnarray}
Finally, we get the following asymptotic behavior for the effective action
for $D=0$ and $R,\lambda <0$: 
\begin{equation}
S\sim \lambda _{L}^{2}\sim \lambda ^{2}r^{4(1-\alpha )h_{-}}  \label{SZDLneg}
\end{equation}
with $h_{-}\equiv C_{1}^{-}/C_{2}^{-}$. Note, that $h_{-}\leq h_{+}$.

Remarkably, the numerical study of the action shows that this scaling found
analytically for the $D=0,R<0$ line holds everywhere in the $Q,R$ plane
except for the Vieillefosse tail, $D=0,R>0$. Hence the main body of the PDF,
which determines the low moments, scales according to (\ref{SZDLneg}). In
order to impose constant energy flux we fix the scaling of the low moments
(including the 3d) to the K41 value, which requires 
\begin{equation}
\alpha =1-\frac{1}{3h_{-}},  \label{Kol}
\end{equation}
thus relating $\alpha $ and $\beta $ parameters of the model. $\alpha $
decreases with $\beta $ increase from $1/3$ at $\beta =0$ (an exceptional
case when tail and the body of the PDF scale the same way) to $0$ at $\beta
\approx 0.3$. We keep $\beta $ as free parameter in the interval $[0,$ $%
\approx 0.3]$ and calculate the respective values  of other exponents numerically
Table 1.

\newpage
\begin{figure}[htbp]
\begin{center}
\setlength{\epsfxsize}{130mm}
\leavevmode
\epsffile{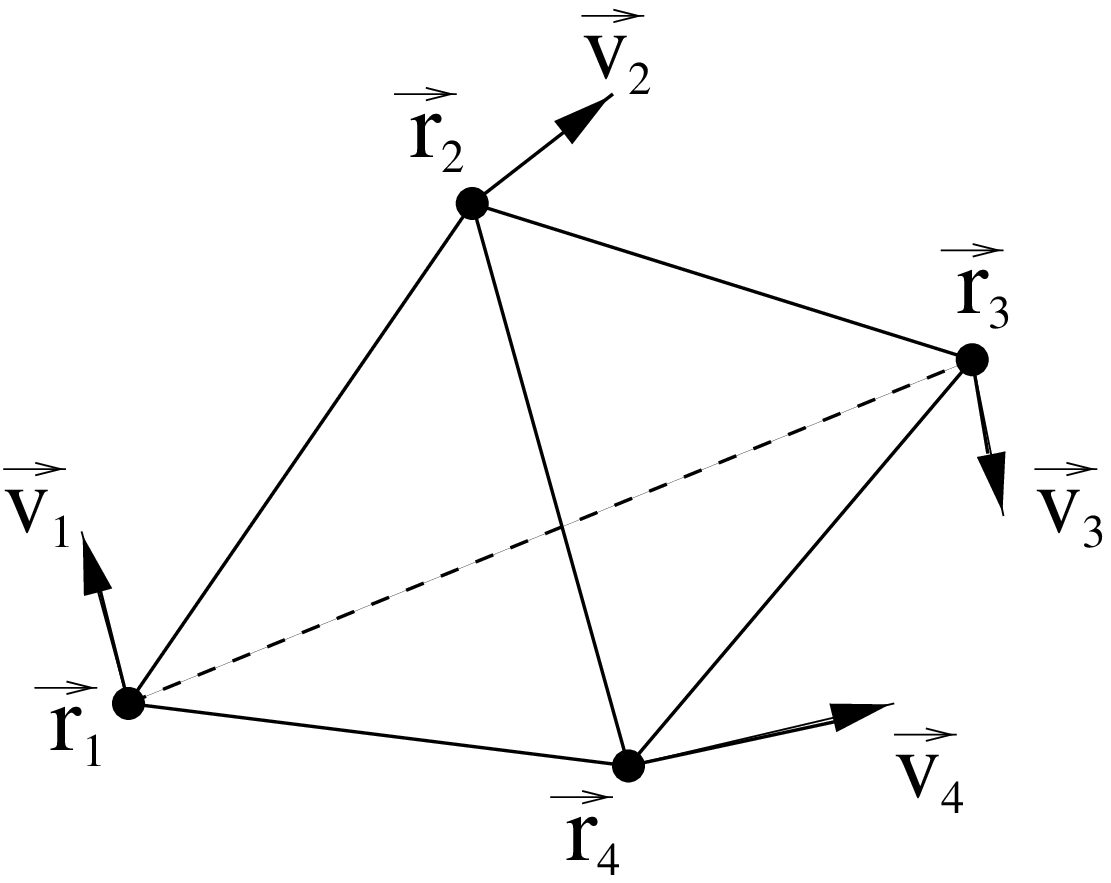}
\vspace{1cm}
\caption{Four Lagrangian points forming a "tetrad". The velocities at
the four points define the coarse grained velocity gradient field.}
\end{center}
\end{figure}

\newpage
\begin{figure}[htbp]
\begin{center}
\setlength{\epsfxsize}{130mm}
\leavevmode
\epsffile{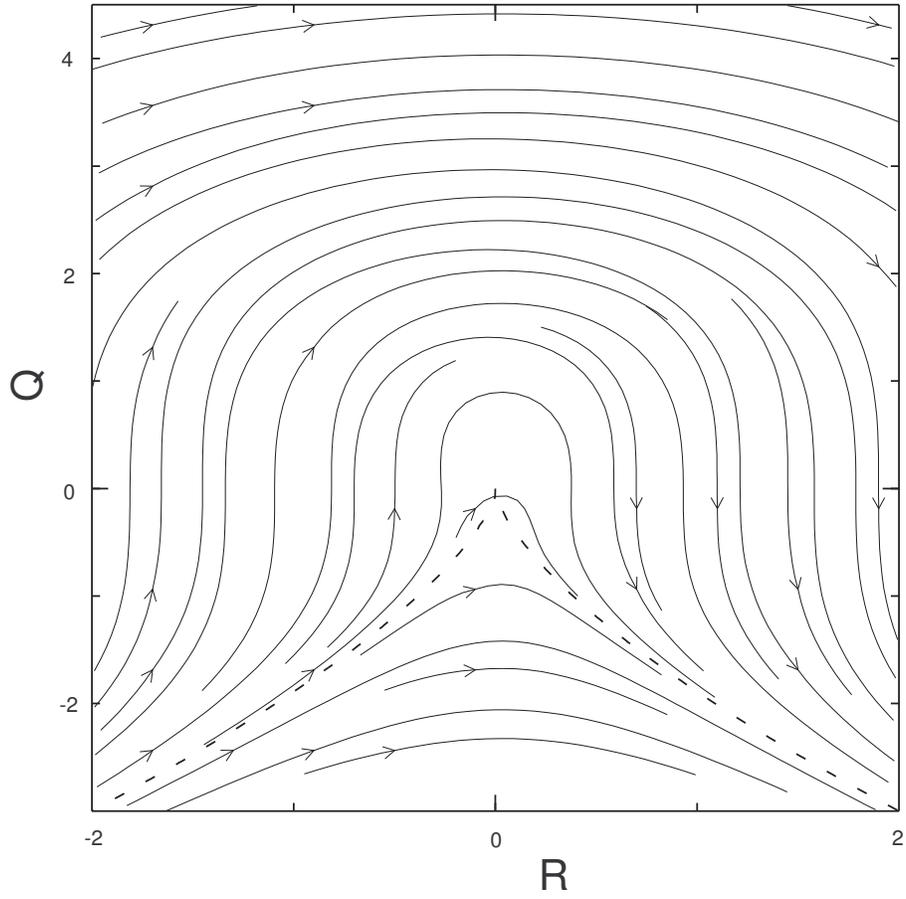}
\vspace{1cm}
\caption{The Restricted Euler flow of velocity gradient invariants $%
Q\,=\, - {\frac{1 }{2}} tr {\bf M}^2$ and $R\,=\, - {\frac{1 }{3}} tr {\bf M}%
^3$.} 
\end{center}
\end{figure}

 

\begin{figure}[htbp]
\begin{center}
\setlength{\epsfxsize}{75mm}
\leavevmode
\epsffile{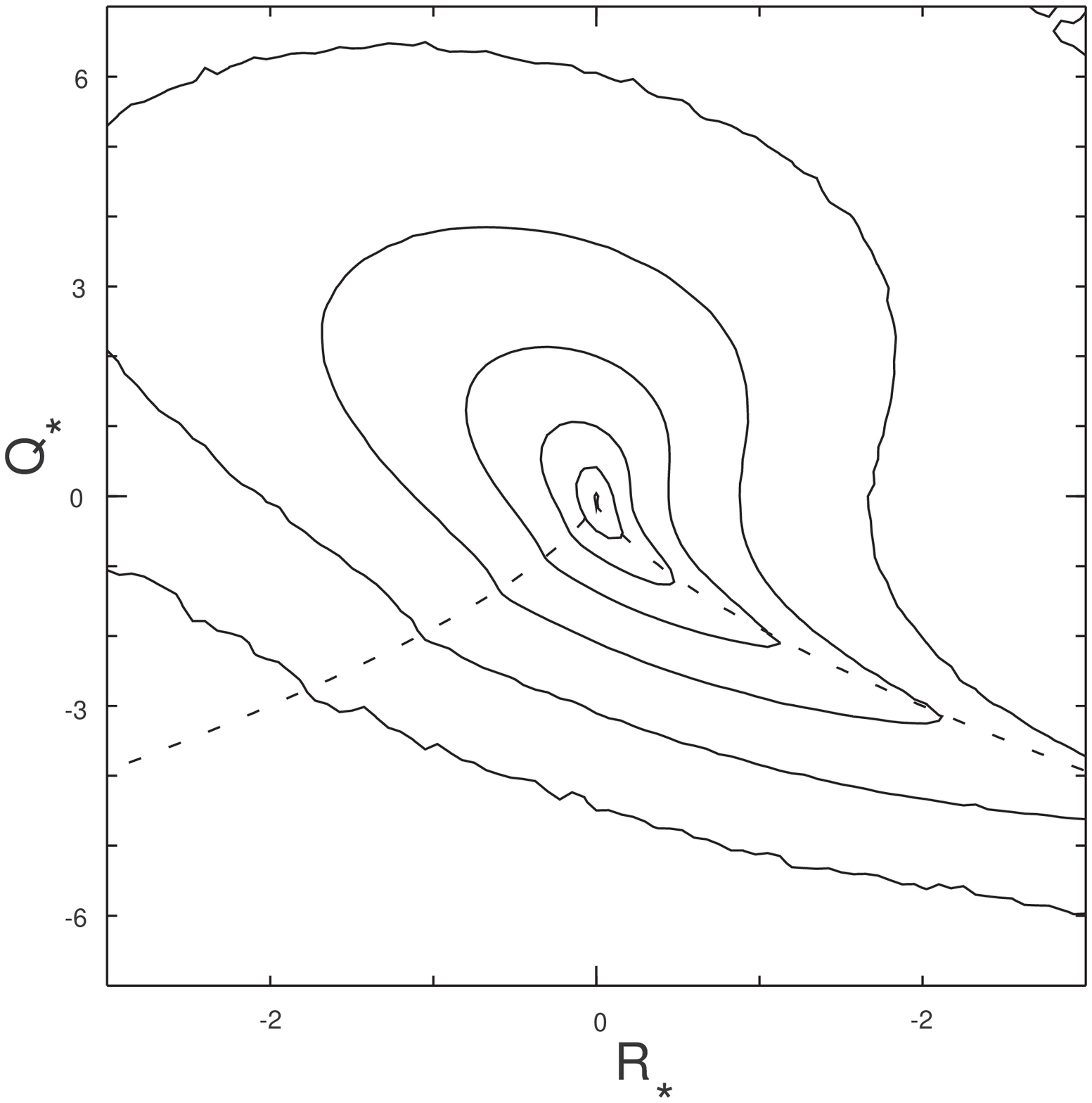 }
\setlength{\epsfxsize}{75mm}
\leavevmode
\epsffile{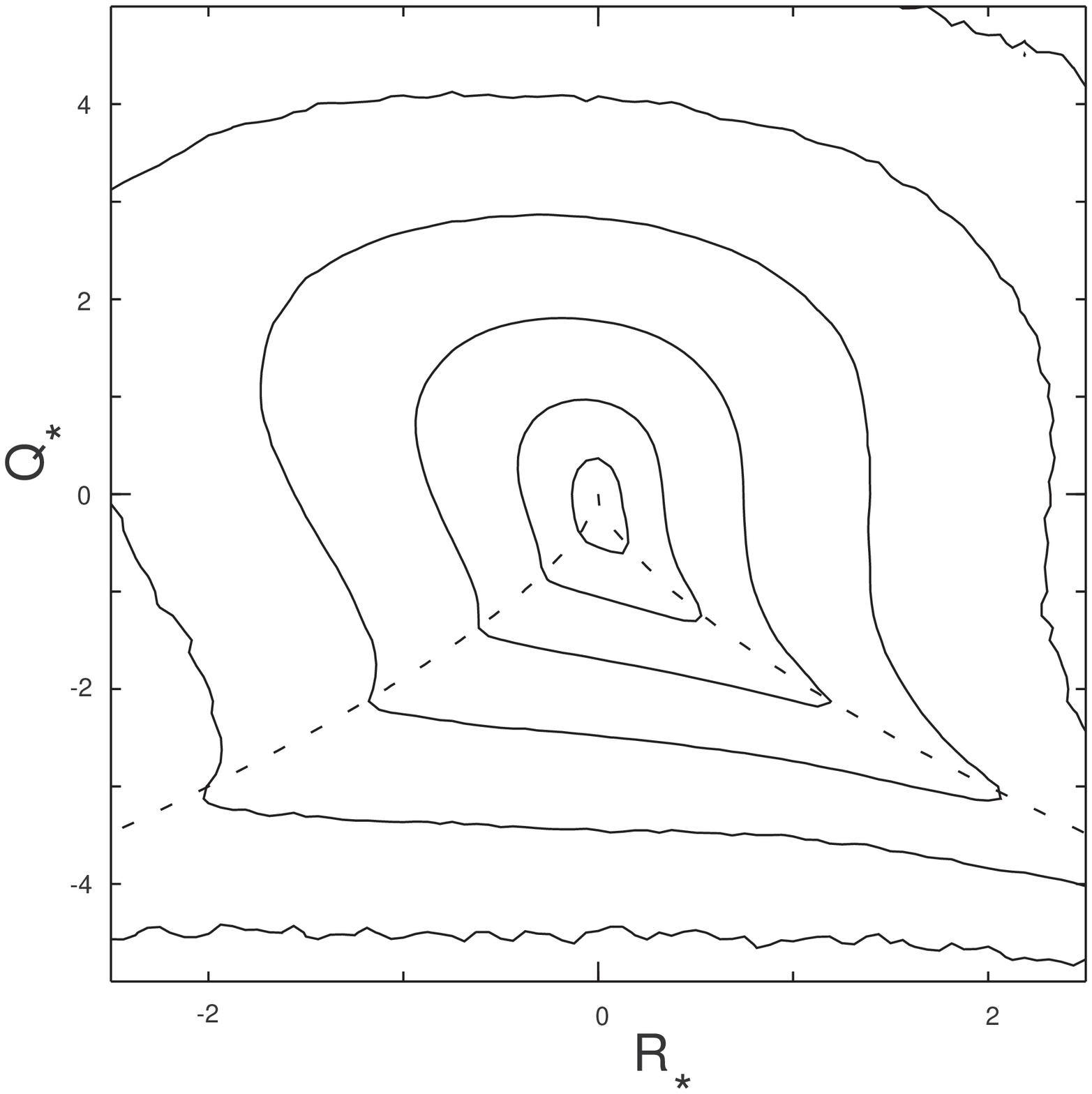 }
\vspace{1cm}
\setlength{\epsfxsize}{75mm}
\leavevmode
\epsffile{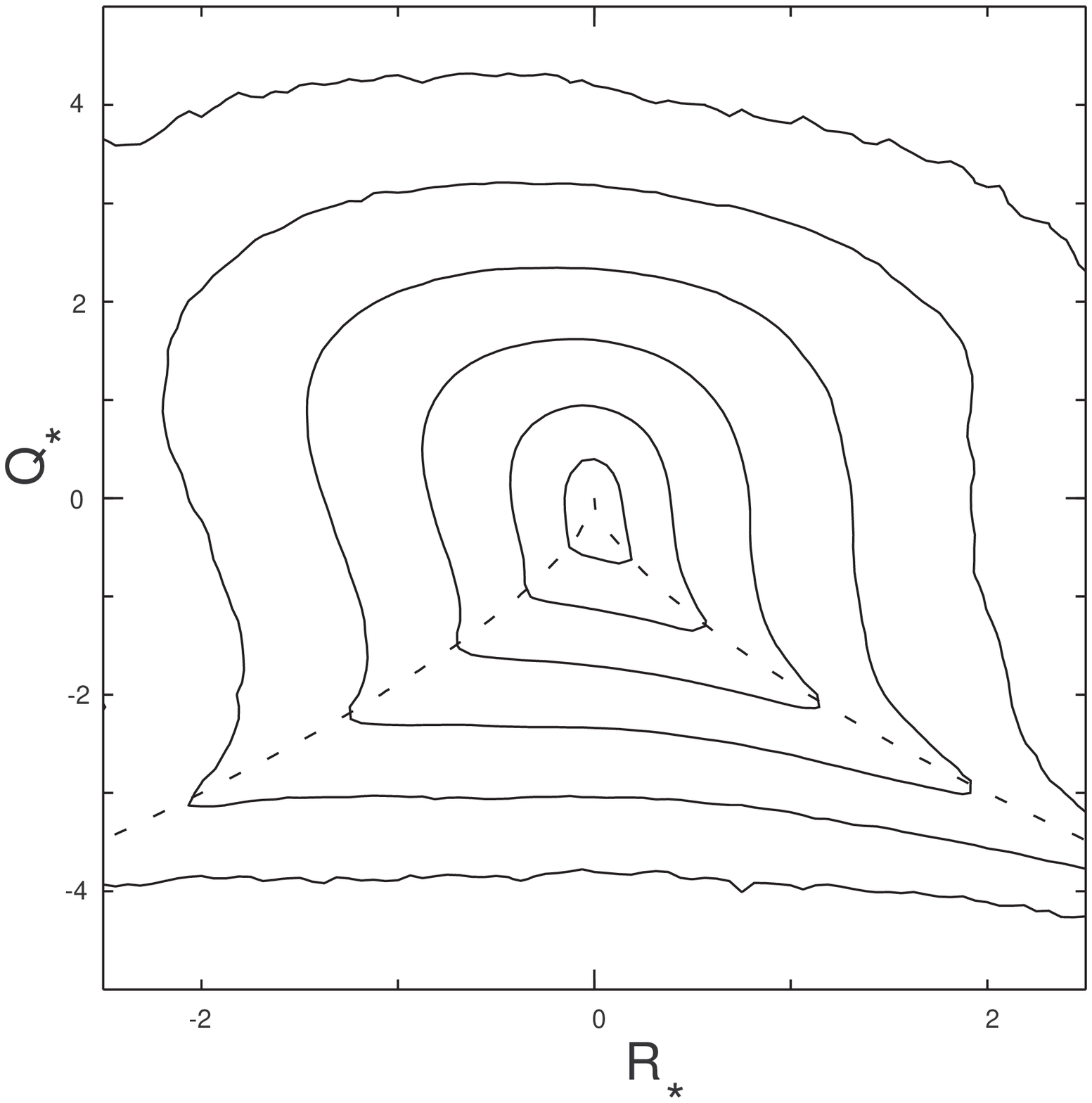 }
\vspace{1cm}
\caption{The PDF of $Q_*,R_*$ invariants normalized to the variance of strain, 
$Q_* \equiv Q/<s^2>$ and $R_* \equiv R/<s^2>^{3/2}$ ("star" denotes normalization), 
obtained from DNS at $R_{\lambda }=85$
measured at different length scales: a) dissipation range $\rho =2\eta $ b)
low end of the inertial range $\rho =8\eta $, and c) upper end of the
inertial range $\rho =L/2$. The isoprobability contours are logarithmically
spaced, and are separated by factors of $10$. The dashed line corresponds to
zero discriminant.}
\end{center}
\end{figure}
\begin{figure}[htbp]
\begin{center}
\setlength{\epsfxsize}{90mm}
\leavevmode
\epsffile{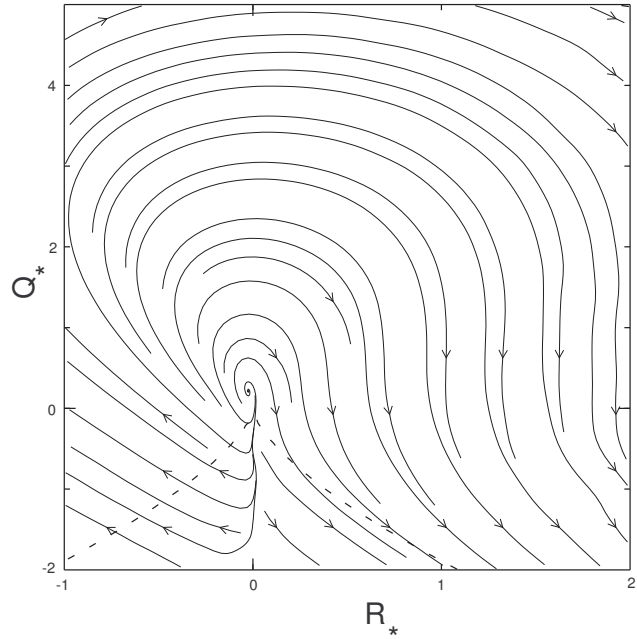 }
\vspace{1cm}
\setlength{\epsfxsize}{90mm}
\leavevmode
\epsffile{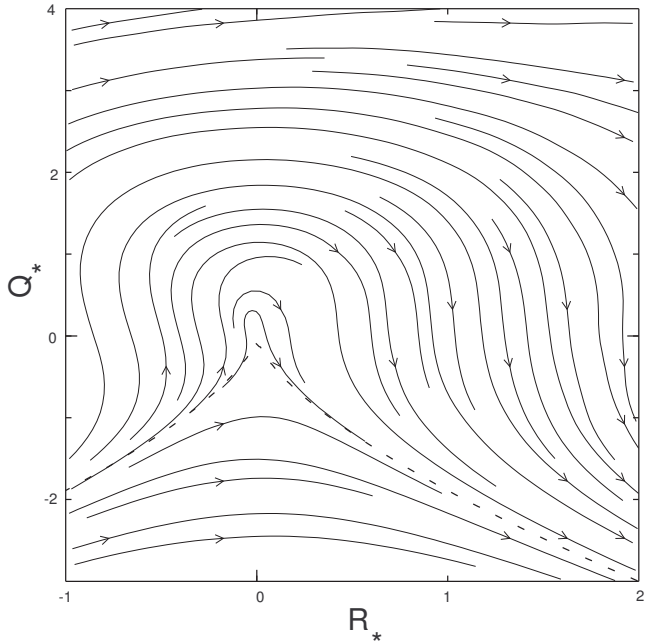 }
\vspace{1cm}
\caption{Streamlines of the flow of $M$-invariants, constructed from the
conditional $<{\dot{Q}}|Q,R>$ and $<{\dot{R}}|Q,R>$ measured from DNS at
different length scales: a) dissipative range $\rho =2\eta =L/32,$ b) upper
end of the inertial range $\rho =L/2$. The invariants $R$ and $Q$ are
normalized as in Fig.3.}
\end{center}
\end{figure}

\begin{figure}[htbp]
\begin{center}
\setlength{\epsfxsize}{90mm}
\leavevmode
\epsffile{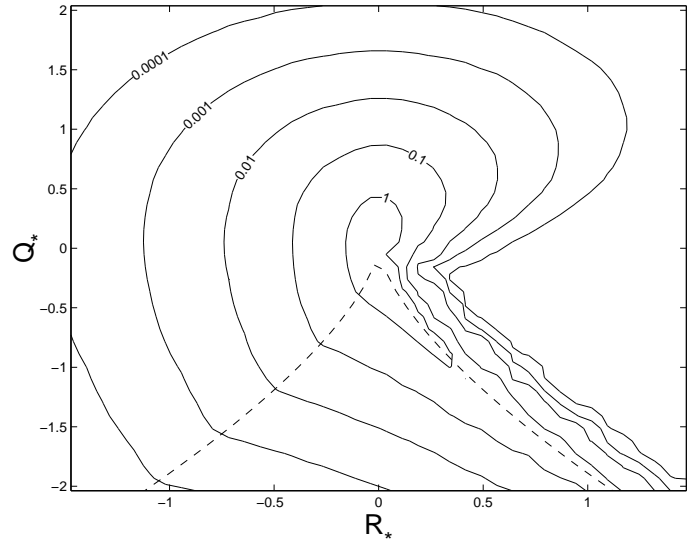}
\vspace{1cm}
\setlength{\epsfxsize}{90mm}
\leavevmode
\epsffile{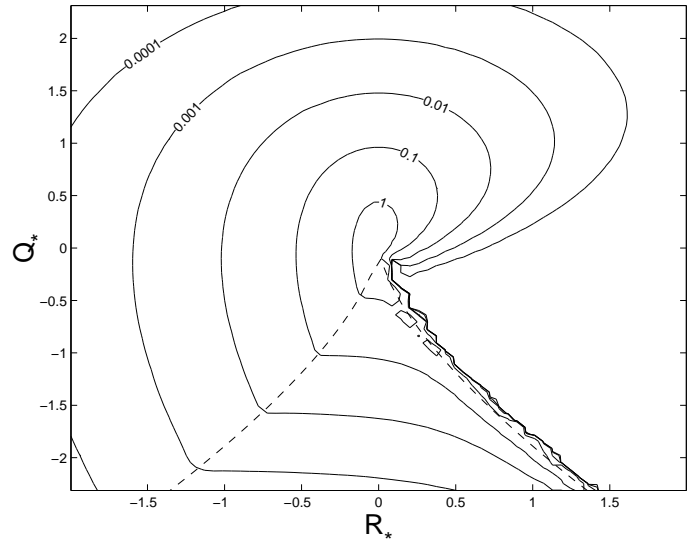 }
\vspace{1cm}
\caption{PDF of $Q_*,R_*$ invariants (normalized as in Fig.3) calculated for the tetrad model in the
deterministic approximation a) $\rho /L=.2$; b) $\rho /L=.5$.}
\end{center}
\end{figure}

\begin{figure}[htbp]
\begin{center}
\setlength{\epsfxsize}{130mm}
\leavevmode
\epsffile{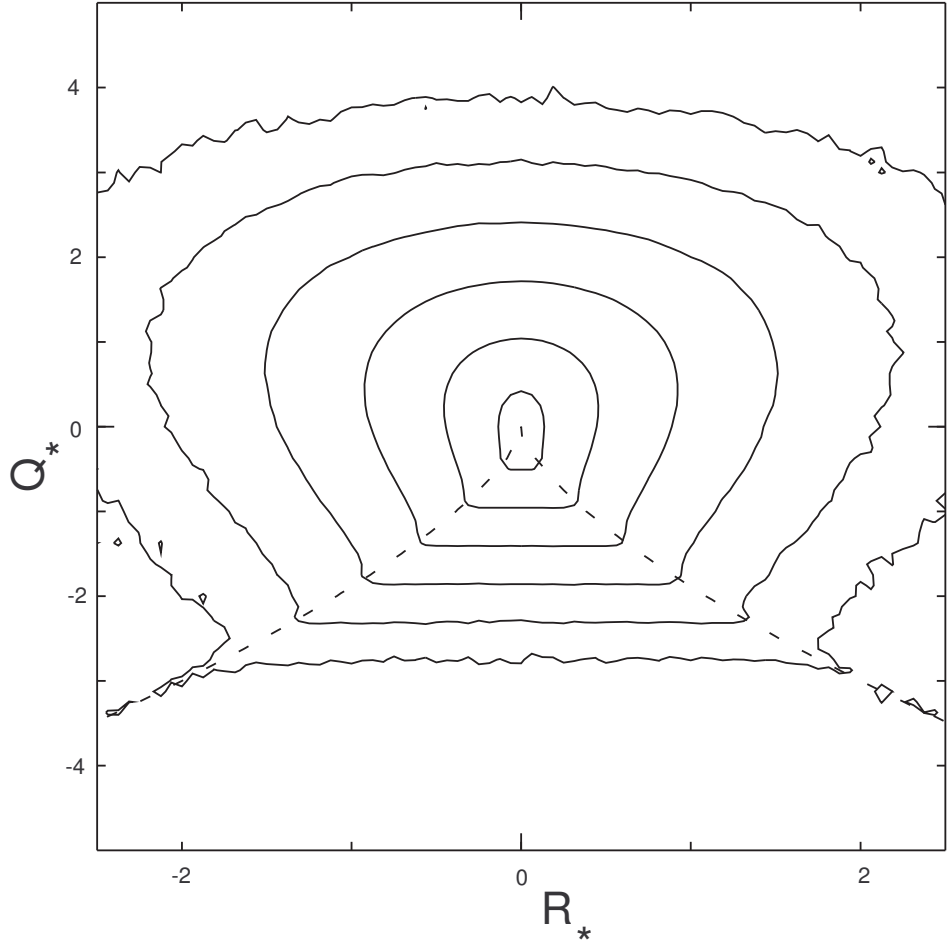}
\vspace{1cm}
\caption{The PDF of $Q_*,R_*$ invariants (normalized as in Fig.3)obtained by replacing the real (DNS)
velocity field by a random Gaussian field, with the same velocity spectrum.}
\end{center}
\end{figure}
\begin{figure}[htbp]
\begin{center}
\setlength{\epsfxsize}{130mm}
\leavevmode
\epsffile{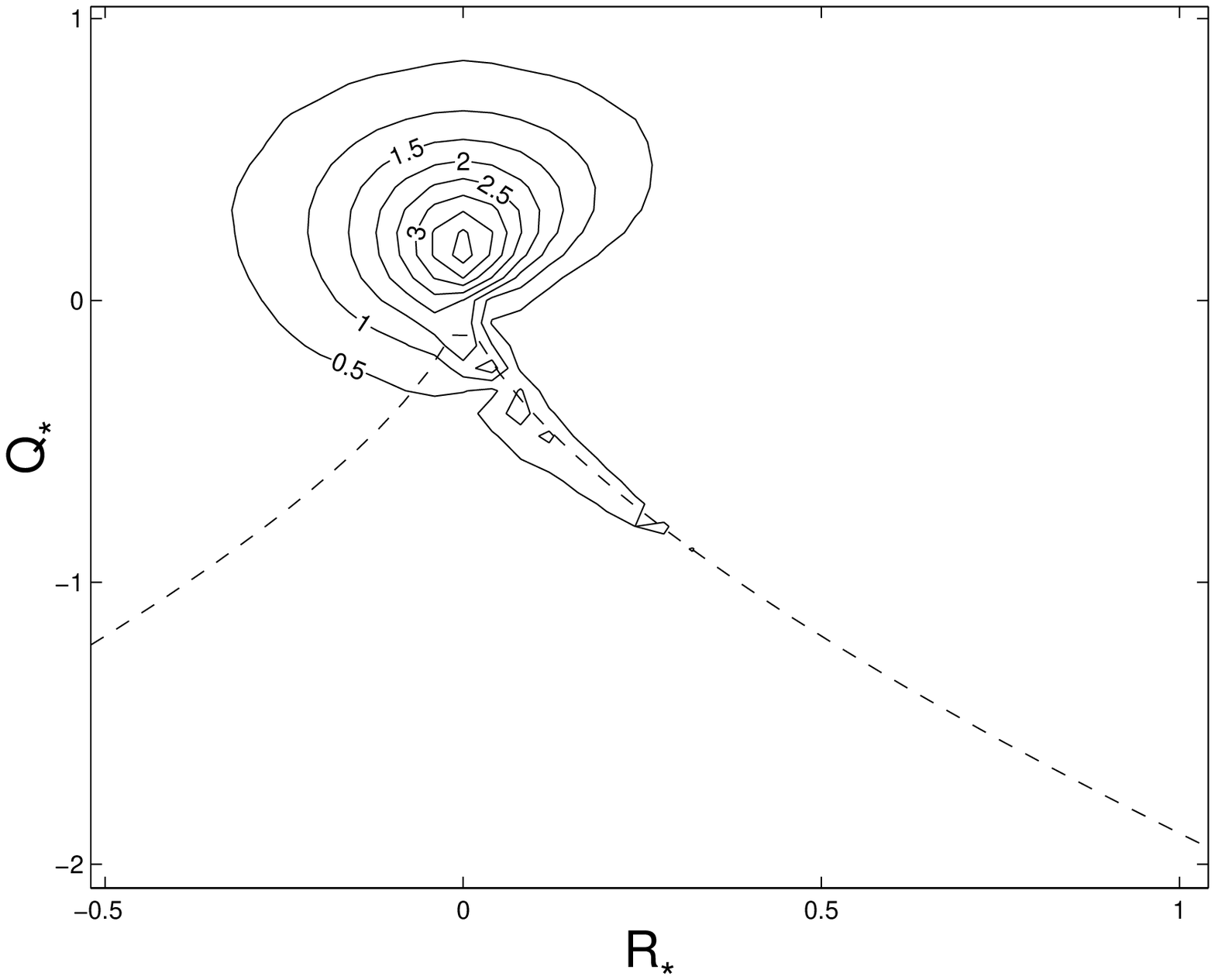}
\vspace{1cm}
\caption{Enstrophy density in $R,Q$ plane for the tetrad model at 
$\rho/L =.5$}
\end{center}
\end{figure}
\begin{figure}[htbp]
\begin{center}
\setlength{\epsfxsize}{130mm}
\leavevmode
\epsffile{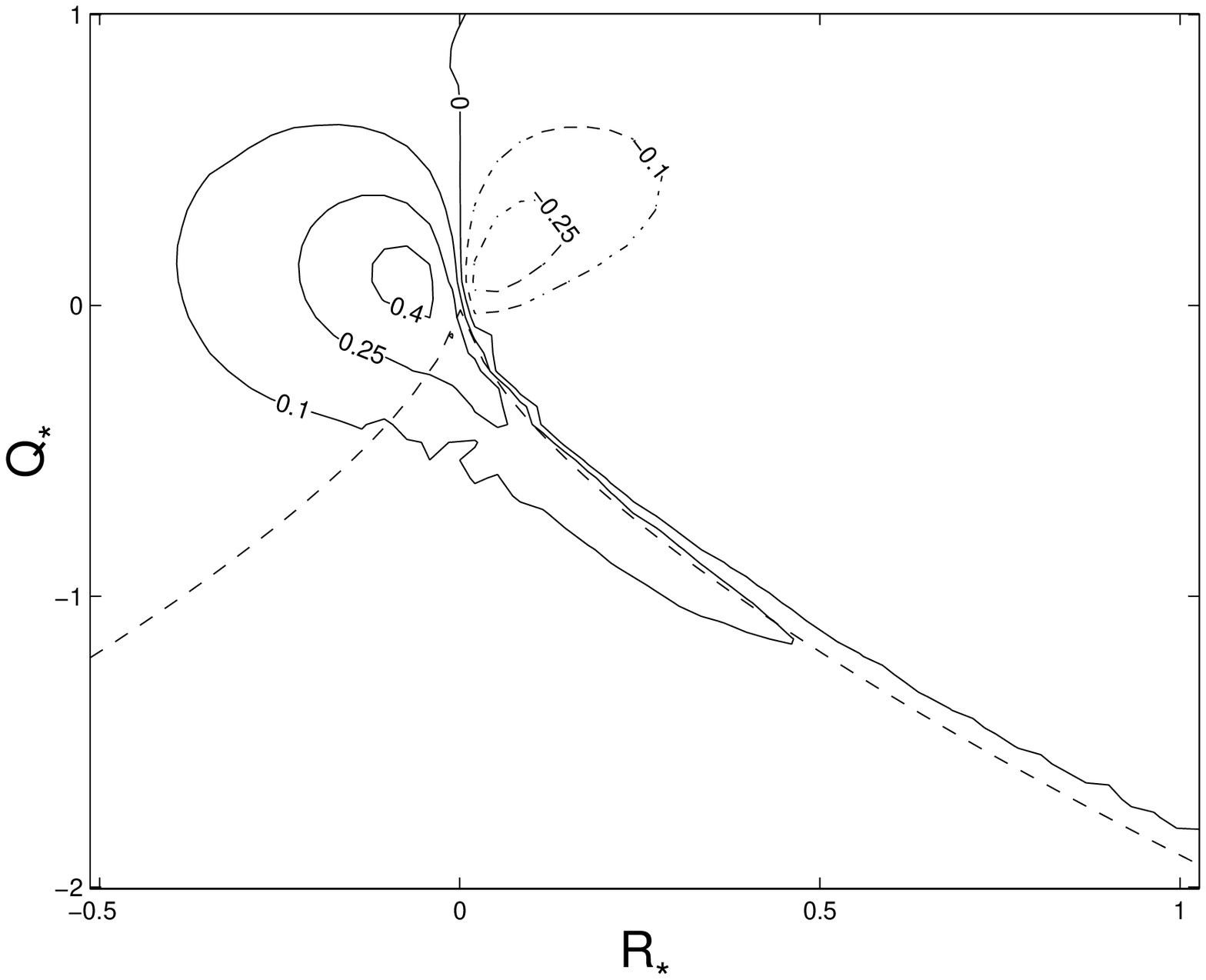}
\vspace{1cm}
\caption{Enstrophy production density in $R,Q$ plane for the tetrad
model at $\rho/L =.5$}
\end{center}
\end{figure}
\begin{figure}[htbp]
\begin{center}
\setlength{\epsfxsize}{130mm}
\leavevmode
\epsffile{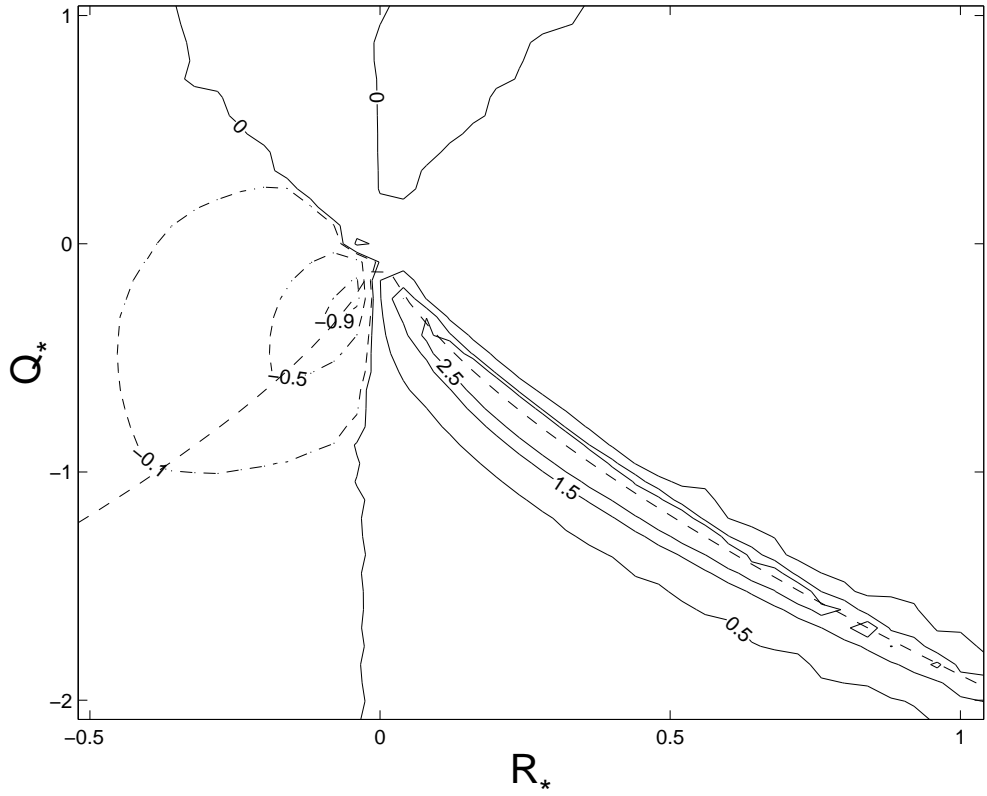}
\vspace{1cm}
\caption{Strain skewness density in $R,Q$ plane for the tetrad model at 
$\rho/L =.5$}
\end{center}
\end{figure}
\begin{figure}[htbp]
\begin{center}
\setlength{\epsfxsize}{130mm}
\leavevmode
\epsffile{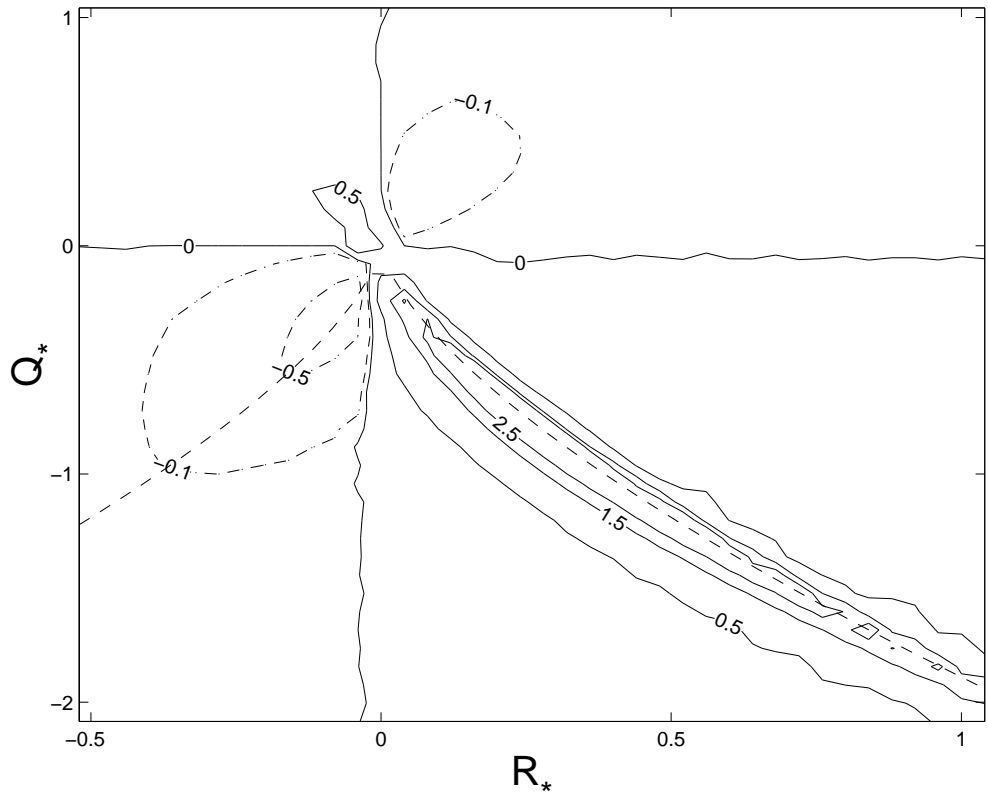}
\vspace{1cm}
\caption{Energy flux density in $R,Q$ plane for the tetrad model at 
$\rho/L =.5$}
\end{center}
\end{figure}
\begin{figure}[htbp]
\begin{center}
\setlength{\epsfxsize}{130mm}
\leavevmode
\epsffile{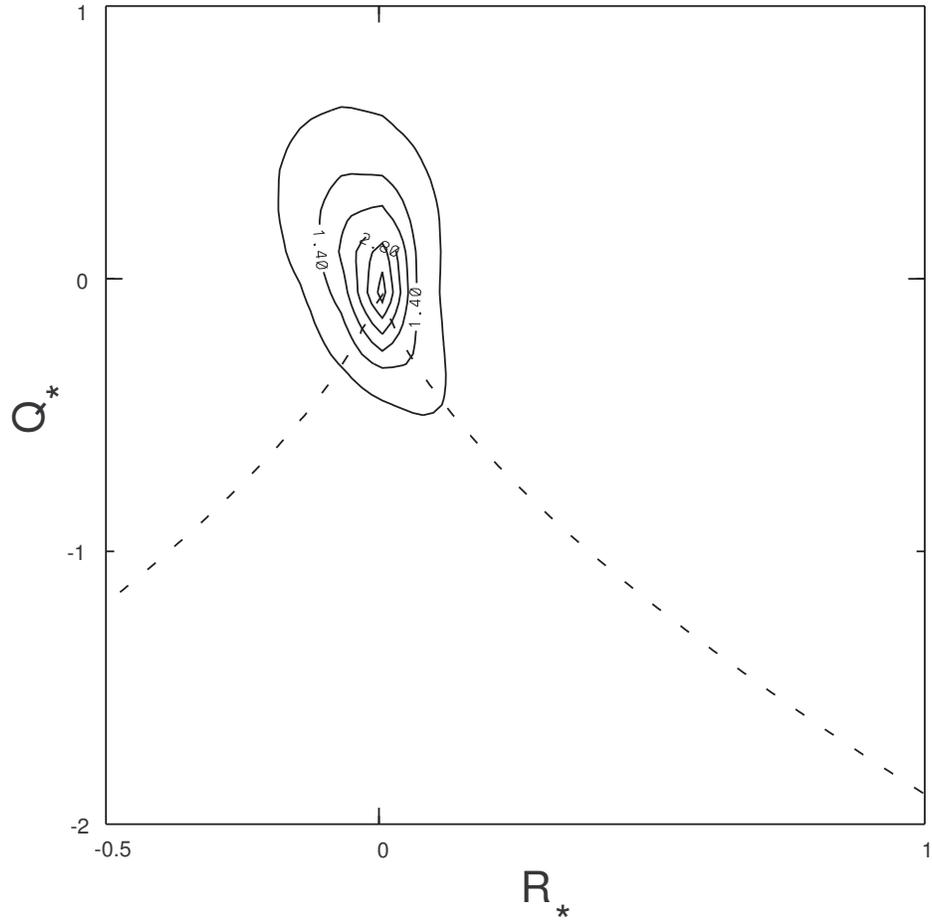}
\vspace{1cm}
\caption{Enstrophy density in $R,Q$ plane from the DNS, $R_{\lambda
}=85 $, at $\rho /L=0.125$ (same value as in Fig.3c,4c). The enstrophy is
normalized by $-2<tr(\Omega ^{2})>$.}
\end{center}
\end{figure}
\begin{figure}[htbp]
\begin{center}
\setlength{\epsfxsize}{150mm}
\leavevmode
\epsffile{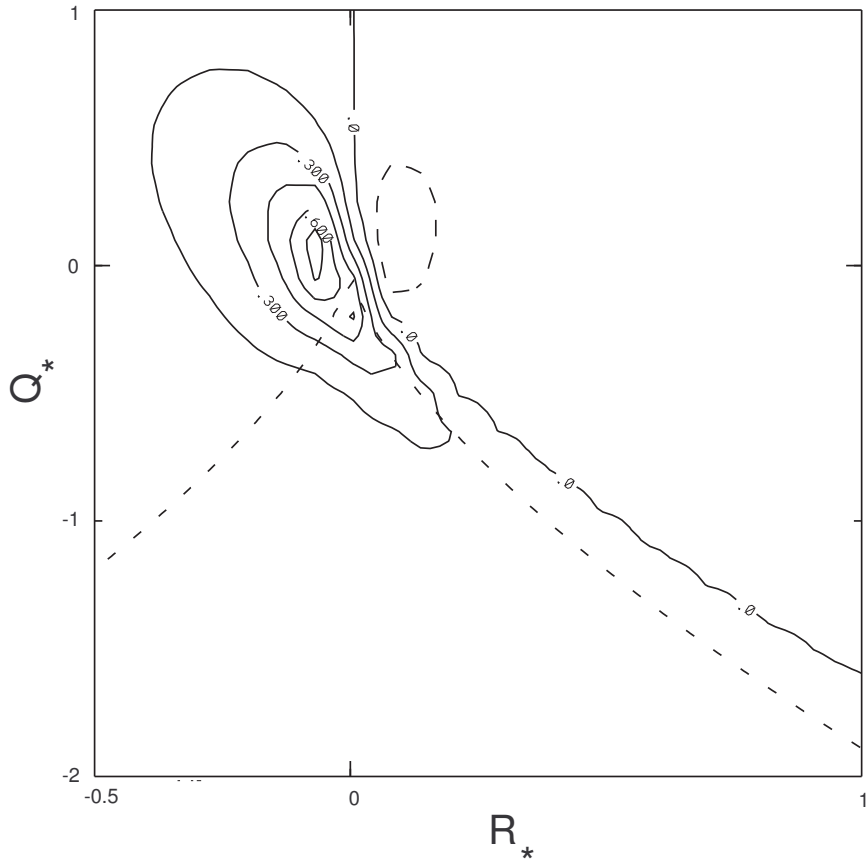}
\vspace{1cm}
\caption{Enstrophy production density in $R,Q$ plane from the DNS, $%
R_{\lambda }=85$, at $\rho /L=0.125$. The enstrophy production is normalized
by $|<tr(M^{2}M^{\dagger })>|$. Solid lines correspond to positive values,
dashed lines to negative values.}
\end{center}
\end{figure}
\begin{figure}[htbp]
\begin{center}
\setlength{\epsfxsize}{130mm}
\leavevmode
\epsffile{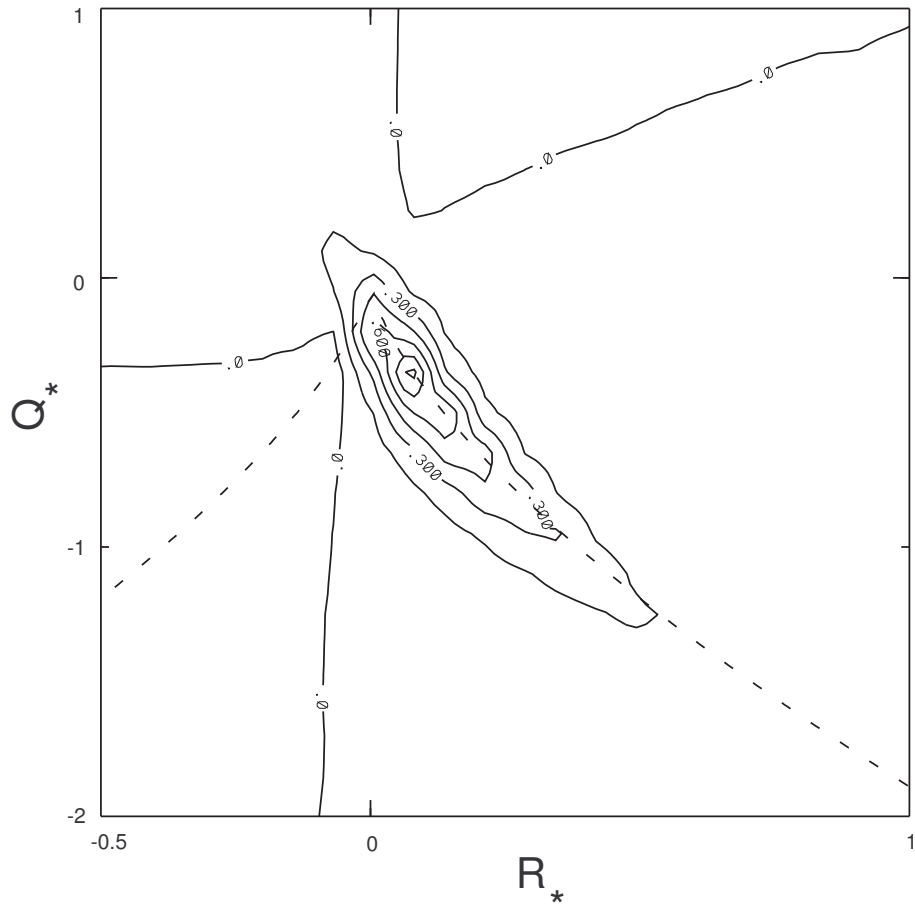}
\vspace{1cm}
\caption{Strain skewness density, $-tr(S^{3})$ in $R,Q$ plane from the
DNS, $R_{\lambda }=85$, at $\rho /L=0.125$. The strain skewness density is
normalized by $|<tr(M^{2}M^{\dagger })>|$. The same convention as in Fig.12
is used.}
\end{center}
\end{figure}
\begin{figure}[htbp]
\begin{center}
\setlength{\epsfxsize}{130mm}
\leavevmode
\epsffile{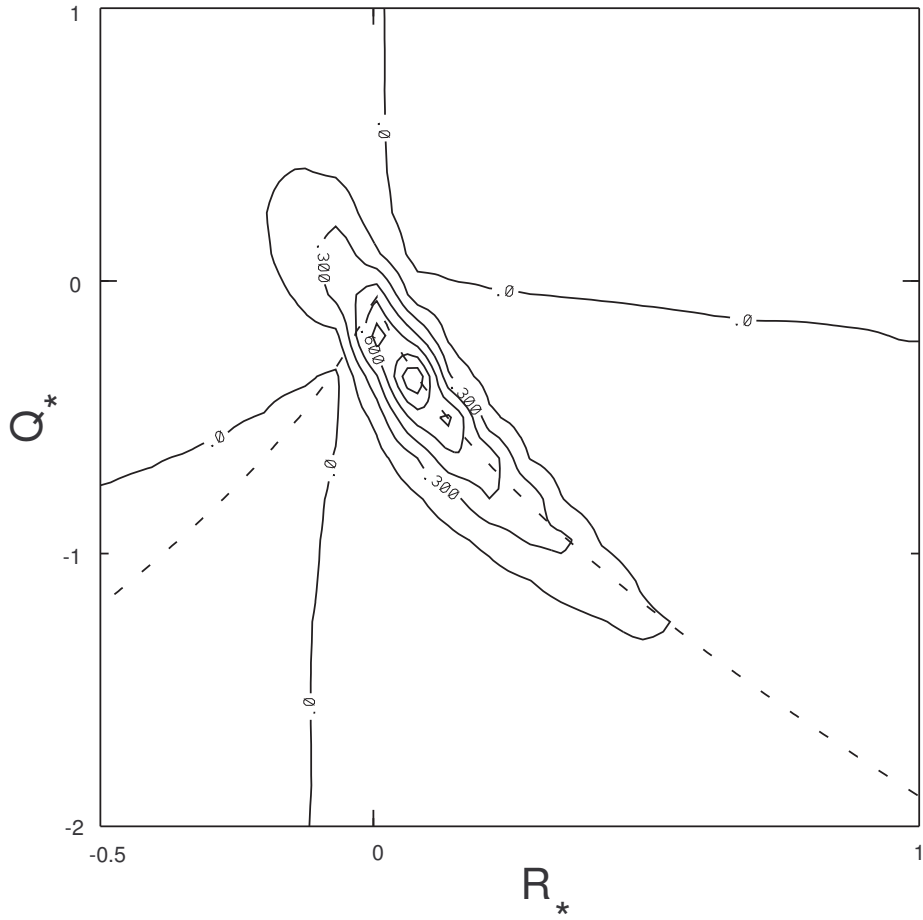}
\vspace{1cm}
\caption{Energy flux density, $tr(M^{2}M^{\dagger })$ in $R,Q$ plane
from the DNS, $R_{\lambda }=85$, at $\rho /L=0.125$. The energy flux density
is normalized by $|<tr(M^{2}M^{\dagger })>|$. The same convention as in
Fig.12 is used.}
\end{center}
\end{figure}
\begin{figure}[htbp]
\begin{center}
\setlength{\epsfxsize}{120mm}
\leavevmode
\epsffile{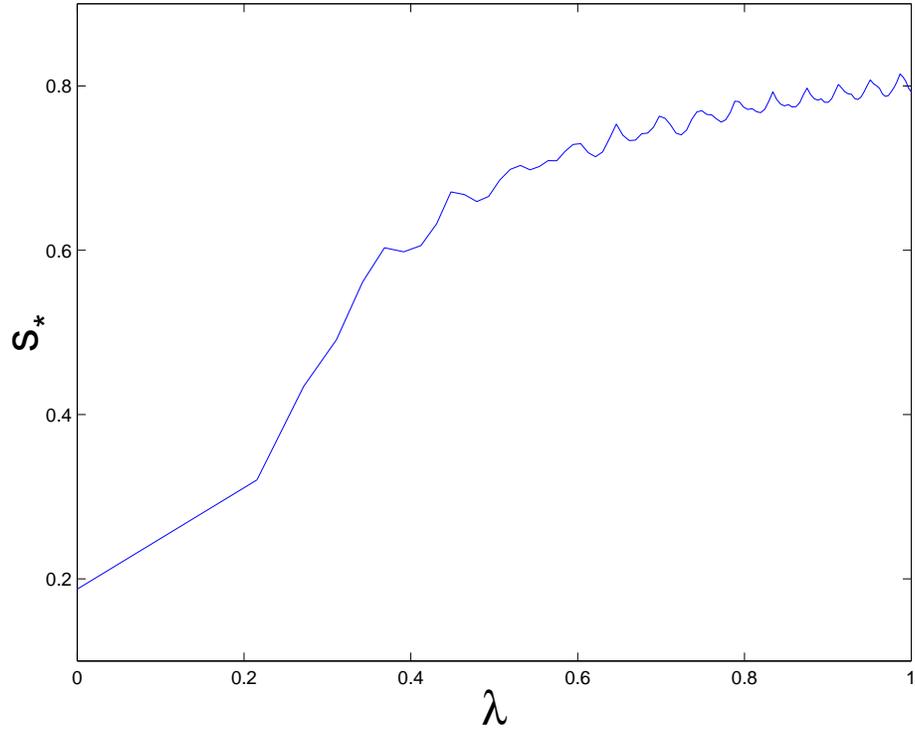}
\vspace{1cm}
\setlength{\epsfxsize}{120mm}
\leavevmode
\epsffile{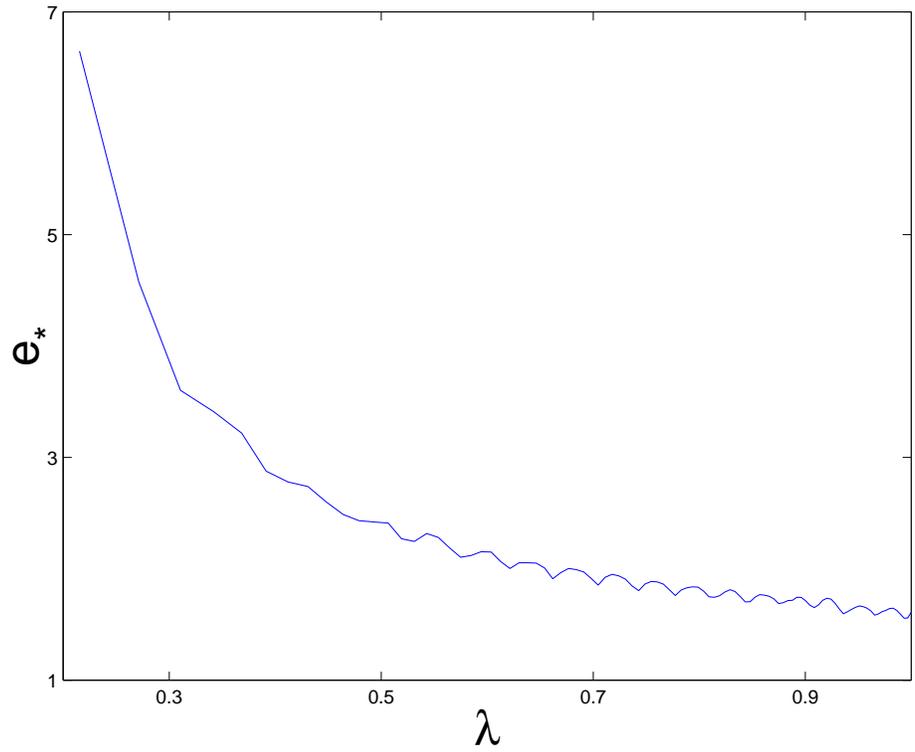}
\vspace{1cm}
\caption{
a) Normalized intermediate eigenvalue of the strain , 
$s_*\equiv[6s_\beta]^{1/2}/|s|$, as a
function of $\lambda $ along the Vieillefosse tail as observed in the DNS, 
$R_\lambda=85$, $\rho =2\eta $; b) same for normalized enstrophy,  
$e_*\equiv\langle\omega^2|\lambda\rangle/\lambda^2$, as a function of $\lambda $.}
\end{center}
\end{figure}

\newpage
$$
\begin{array}{lllllll}
\beta & 0.05 & 0.1 & 0.15 & 0.2 & 0.25 & 0.3 \\ 
\alpha & 0.29 & 0.24 & 0.19 & 0.14 & 0.07 & 0.005 \\ 
\gamma _{+} & 0.05 & 0.11 & 0.18 & 0.25 & 0.33 & 0.43 \\ 
\eta & 1.11 & 1.14 & 1.21 & 1.28 & 1.35 & 1.44 \\ 
\zeta & 1.95 & 1.86 & 1.71 & 1.66 & 1.63 & 1.55 \\ 
\delta & 1.33 & 1.69 & 1.69 & 1.85 & 1.89 & 1.90 \\ 
\eta ^{\prime } & 0.27 & 0.22 & 0.34 & 0.34 & 0.42 & 0.52 \\ 
n_{c} & 62.98 & 28.87 & 18.82 & 12.88 & 9.92 & 7.52
\end{array}
$$

{\bf  Table 1} Table of exponents for the $D=0$, $R>0$ tail  in the deterministic
limit (Appendix A).


\begin{references}
\bibitem{ft1} This expectation follows from examining the exact expression for the
pressure Hessian in terms of the integral over $ tr (\partial v)^2$ and 
observing that upon
coarse-graining over region of size $R$ it reduces approximately to a sum over many uncorrelated
volumes (of size $R$). For large "d" the number of uncorrelated 
R-volumes in a shell surrounding the  central - local - region is large.
\bibitem{ft2} The relation of pressure anisotropy $\Pi_{a,b}$ to the shape of the tetrad
given by (3c) is presented here as a model assumption. Alternatively, one could
start with the pressure anisotropy tensor (defined over spherical volume of size $R$) and
study its Lagrangian dynamics which, if the present assumption is correct,
should be consistent with (3b,c).
\bibitem{ft3} The Lagrangian path representation also allows incorporation of
viscosity (in analogy with scalar diffusivity\cite{SS94}) via addition of a Langevin
noise source in (3b) or equivalently an additional $\nu \delta_{a,b} \delta_{i,j} \delta (t)$
term in (5b). It however plays no role as long as one works with pure inertial range,
coarse grained quantities.
\bibitem{ft5} This sort of singular solutions have been
earlier noted by others (e.g. Childress and Siggia - private communications).
\bibitem{ft6} We have already
seen that the introduction of the ${\bf \Pi }$ tensor also eliminates the
unphysical contribution of the local pressure to $\partial _{t}V^{2}$ which
underlies the singularity \cite{VFS}.
\bibitem{ft7} By ribbon we mean configuration with two unequal stretching directions.
\bibitem{ft8} For simplicity we shall for the time being ignore the physical constraint
imposed on the PDF by the homogeneity $<tr\,{\bf M}^{2}>=0$. It can be
easily added.
\bibitem{ft9}This is not an essential approximation
and proper integration may be performed numerically.
\bibitem{ft11} The degree of alignment of course would vary if
conditioned on $\sigma $ or $e$ as an independent variable.
\bibitem{ft13} While the question of
existence of finite times singularities is still being debated, it is clear
that it has to be generated by some non-local effect, rather than by local
stretching effect \cite{PS90,PS92,Kerrsing,Pelz,Grauer}.
\bibitem{ft16}  Note that the "tails" here refer to the intermediate asymptotics
where $\lambda$ is much larger than its r.m.s. value but smaller
than the characteristic large scale velocity over $r$. The latter
condition is imposed by the bound on velocity difference PDF imposed
by the single point PDF \cite{Noullez}.
\end{references}
\end{document}